\documentclass[12pt,a4paper]{article}
\pdfoutput=1

\usepackage{amsmath,amssymb,bm,cite,changes}

\usepackage{amsmath}
\usepackage{amssymb,xfrac,mathrsfs}
\usepackage{graphicx}
\usepackage{caption}
\usepackage{subcaption}
\usepackage{bm}% bold math
\usepackage[bbgreekl]{mathbbol}
\usepackage{cite,color,float}
\usepackage{enumitem}
\usepackage{color}
\usepackage{comment}
\usepackage[toc,page]{appendix}
\usepackage{tikz}
\usepackage{slashed}
\usetikzlibrary{decorations.markings}
\usepackage{manfnt}
\usepackage{relsize}

\newcommand\shrink{\hspace{.4em}}

\newcommand\undermat[2]{%
  \makebox[0pt][l]{$\smash{\underbrace{\phantom{%
    \begin{matrix}#2~\end{matrix}}}_{\text{$#1$}}}$}#2}

\def\msp{\!\mathsmaller{\mathsmaller{\bm{+}}}}
\def\msm{\!\mathsmaller{\mathsmaller{\bm{-}}}}
\def\msz{\!\mathsmaller{\mathsmaller{\bm{0}}}}

\def\qmb{\bm{q}\!\cdot\!\bm{M}}
\def\qmpb{\bm{q'}\!\!\cdot\!\bm{M}}

\def\vred{\displaystyle{\widehat{\mathop{V}_-}}}%_{\!\vee}}
\def\rd{\mathrm{d}}
\def\ri{\mathrm{i}}
\def\I{\mathrm{I}}

\def\np{N_\mathrm{P}}
\def\nl{N_\mathrm{L}}
\def\nd{N_\mathrm{d}}

\tikzset{    arrow/.style={decoration={markings, mark=at position 1 with
    {\fill(-0.09*#1,-0.03*#1) -- (0,0) -- (-0.09*#1,0.03*#1) -- cycle;}}, postaction={decorate}},
    arrow/.default=1}

\def\linkcur{\tikz[baseline=-.6ex]{\draw[arrow=1.8] (-.5,0) -- (0.1,0);\draw[-] (.1,0) -- (.5,0);}}

\def\linkcurinv{\tikz[baseline=-.6ex]{\draw[-] (-.5,0) -- (0.1,0);\draw[arrow=1.8] (.5,0) -- (-.1,0);}}

\def\linkdash{\tikz[baseline=-.6ex]{\draw[arrow=1.8,dashed] (-.5,0) -- (0.1,0);\draw[dashed] (.1,0) -- (.5,0);}}

\def\linkdashinv{\tikz[baseline=-.6ex]{\draw[dashed] (-0.1,0) -- (-.5,0);\draw[arrow=1.8,dashed] (.5,0) -- (-.1,0);}}

\def\circ{\tikz[baseline=-.5ex]{
		\draw [dashed] circle (.412); \draw[arrow=1.8] (.32,.25) -- (.3,.28); \draw[arrow=1.8] (-.32,-.25) -- (-.3,-.28);}}
	
\def\circinv{\tikz[baseline=-.5ex]{
		\draw [dashed] circle (.412); \draw[arrow=1.8](.3,.28)  -- (.33,.25); \draw[arrow=1.8] (-.3,-.28) -- (-.33,-.25);}}

\def\recdashcirc{\tikz[baseline=-.5ex]{\draw (-.55,-.55) rectangle (.55,.55);
\draw[arrow=1.8] (.1,.55) -- (-.1,.55); \draw[arrow=1.8] (-.1,-.55) -- (.1,-.55);
\draw[arrow=1.8] (.55,-.1) -- (.55,.1); \draw[arrow=1.8] (-.55,.1) -- (-.55,-.1);
\draw [dashed] circle (.412); \draw[arrow=1.8] (.3,.28) -- (.33,.25); \draw[arrow=1.8] (-.3,-.28) -- (-.33,-.25);}}

\def\recdashrec{\tikz[baseline=-.5ex]{\draw (-.55,-.55) rectangle (.55,.55);
\draw[arrow=1.8] (.1,.55) -- (-.1,.55); \draw[arrow=1.8] (-.1,-.55) -- (.1,-.55);
\draw[arrow=1.8] (.55,-.1) -- (.55,.1); \draw[arrow=1.8] (-.55,.1) -- (-.55,-.1);
\draw (-.4,-.4) [dashed] rectangle (.4,.4);
\draw[arrow=1.8] (-.1,.4) -- (.1,.4); \draw[arrow=1.8] (.1,-.4) -- (-.1,-.4);
\draw[arrow=1.8] (.4,.1) -- (.4,-.1); \draw[arrow=1.8] (-.4,-.1) -- (-.4,.1);}}

\def\twocirc{\tikz[baseline=-.5ex]{
\draw [dashed] circle (.412); \draw[arrow=1.8] (.3,.28) -- (.33,.25); \draw[arrow=1.8] (-.3,-.28) -- (-.33,-.25);
\draw [dashed] circle (.288); \draw[arrow=1.8] (.29,.05) -- (.21,.2); \draw[arrow=1.8] (-.29,-.05) -- (-.21,-.2);}}

\def\backforth{\tikz[baseline=-.6ex]{\draw[arrow=1.8,dashed] (0,-.5) -- (0,0);\draw [-,dashed] (0,.1) -- (0,.5);
\draw[arrow=1.8,dashed] (.1,.42) -- (.1,-.1);\draw [-,dashed] (.1,-.2) -- (.1,-.5);}}

\begin{document}

~
\vskip 1cm 

\begin{center}
\noindent {\large\textbf{Weak Coupling Limit of U(1) Lattice Model \\
in Fourier Basis
}}
%\maketitle \thispagestyle{empty}
\\[1.5\baselineskip]
Afsaneh Kianfar%~\footnote{} 
~~~~~and~~~~~ 
Amir H. Fatollahi~\footnote{Corresponding Author: fath@alzahra.ac.ir}
\\[1.5\baselineskip]
\textit{Department of Physics, Faculty of Physics and Chemistry, \\
Alzahra University, Tehran 1993891167, Iran}
\\[2\baselineskip]
\begin{abstract}

\noindent
The transfer-matrix of the U(1) lattice model is considered in the Fourier basis and in
the weak coupling limit.
The issues of Gauss law constraint and gauge invariant states are addressed in the 
Fourier basis. In particular, it is shown that in the strong coupling limit
the gauge invariant Fourier states are effectively the finite size closed loop currents. 
In the weak coupling limit, however, the link-currents along periodic or infinite spatial 
directions find comparable roles as gauge invariant states. 
The subtleties related to the extreme weak coupling of the transfer-matrix in the
Fourier basis are discussed. A careful analysis of the zero 
eigenvalues of the matrix in the quadratic action leads to a safe extraction 
of the diverging group volume in the limit $g\to 0$. 
By means of the very basic notions and tools of the lattice model, the spectrum at
the weak coupling limit for any dimension and size of lattice is obtained analytically. 
The spectrum at the weak coupling limit corresponds to the expected one 
by the continuum model in the large lattice limit.
\end{abstract}
\end{center}

\vskip 2cm


 \newpage  

\section{Introduction and Outline}
In spite of the incomparable advantages of lattice gauge models 
for theoretical and practical purposes in the strong coupling regime, 
a full understanding of the weak coupling limit is still an open issue. 
In particular, the lattice gauge models are usually transferred 
to the extreme weak (arbitrary small) coupling regime 
in an uncontrolled way, leaving shortcomings and unresolved issues, specifically as follows:
\begin{enumerate}
\item Although the distinguished role of the Wilson loops is well appreciated in the strong coupling limit, 
it is not quite understood which subset of these gauge invariant quantities has the main role in 
the weak coupling regime.
\item In going to the extreme weak coupling limit, the exact point at which the diverging contribution of 
the unfixed gauge degrees comes into the calculation is not well identified. Further, a way of safe 
and controlled extraction of the diverging contribution, the so-called group volume, is still lacking. 
\item Once the above two issues are fixed, as a matter of necessity, the lattice model 
has to recover the continuum part of the spectrum expected by the classical model in any dimension. 
\end{enumerate}
It is the purpose of the present work to address the above issues for the pure U(1) lattice gauge 
model in the temporal gauge. In particular, the transfer-matrix of the model is studied in the 
field Fourier basis. 
The lattice gauge models in Fourier basis have been studied since the early 
years of these models. The plaquette degrees in Fourier basis, the so-called 
dual variables \cite{savit}, are used to present a qualitative description of the phases 
by the U(1) model in different dimensions \cite{banks}. 
The numerical studies based on the dual formulation show a clear advantage of using 
integer variables compared to the original continuous ones \cite{gatt}.
The Fourier transform of the links variables also is known as the electric flux basis. 
The basis is used in a numerical setup of the Hamiltonian formulation of the lattice models 
\cite{koonin,byrnes,best}.
Following \cite{vadfat,kiafat}, the present application of the Fourier transform is also 
for gauge link variables appearing in the transfer-matrix of the model. 
The associated Fourier variables again appear to be integer 
valued, and they are interpreted as quantized currents on the lattice links \cite{kiafat}.
In \cite{vadfat,kiafat} it was shown that the transfer-matrix in the Fourier basis is block diagonal.  
In fact, as a consequence of a lattice version of the local current conservation, the current-states 
differing in the loop-currents circulating inside plaquettes belong to the same block \cite{kiafat}. 
The members of each block 
can be constructed by an arbitrary member of the block as the representative \cite{kiafat}. 
A diagrammatic representation was introduced in \cite{kiafat} for the strong coupling expansion 
of the transfer-matrix elements in the Fourier basis. With $g$ as the gauge coupling, 
the parameter of expansion is $1/g^2$, which is small in the strong coupling 
limit. The expansion of the matrix-element between two current-states of the same block is directly interpreted as the 
occurrence of all possible \textit{virtual link and loop currents} that transform the current-states to 
the vacuum (the state with no current). Based on the expansion of the 
transfer-matrix elements in the Fourier basis, it was shown that the low lying energy levels can be calculated by means 
of the simple perturbative methods in the strong coupling regime \cite{kiafat}.

Based on notions and expressions developed in \cite{kiafat}, as far as the transfer-matrix 
of the U(1) gauge model in the Fourier basis is concerned, it is shown that the above mentioned issues 
can be addressed, as the following: 
\begin{enumerate}
\item In the Fourier basis the Wilson loops are represented by the current-states with no 
boundary, which are either current-loops belonging to the vacuum block, or states with 
equal links-currents along a periodic or an infinite spatial direction. While in the strong 
coupling limit the closed currents of the vacuum block have the main contribution to the transfer-matrix, 
in the weak coupling regime the link-currents along spatial directions also find comparable roles. 
\item In the extreme weak coupling limit, the zero eigenvalues of the matrix in the quadratic action 
can be identified as the origin of the diverging contributions to the elements of the transfer-matrix in the 
Fourier basis. The states belonging to the subspace corresponding to the zero eigenvalues are clearly interpreted as 
pure gauge configurations, on which the matrix in the quadratic action vanishes. The dimension, as well as the 
diverging volume of the subspace in the weak coupling limit, can be handled in a safe and controlled way.
\item By means of the very basic notions and tools of the lattice model, the spectrum at
weak coupling limit for any dimension and size of lattice is obtained analytically. 
The spectrum consists of the sum of possible energies by 
static and standing wave field configurations on the lattice.
In the large lattice limit the spectrum at the weak coupling limit corresponds to the expected one 
by the continuum model.
\end{enumerate}

One of the basic tools used in the formulation of the transfer-matrix in the Fourier basis is 
the plaquette-link matrix $\bm{M}$ \cite{vadfat,kiafat}, by which the elements defined on the 
lattice can be managed at any coupling.
As a consequence, it is seen that the matrix $\bm{M}$ 
provides the possibility to keep and work with the fundamental lattice notions 
even in the extreme weak coupling limit. On the other hand, using this matrix enables to 
translate the tools in the continuum model, such as spatial derivatives, into the lattice model.
These all make it possible to calculate the dimension of the subspace by the mentioned 
zero eigenvalues and to control the group volume.  

The rest of the paper is organized as follows. In Sec.~2,
a short review of the formulation of the transfer-matrix in the Fourier basis, together with 
a detailed description of emerging notions, are presented. 
In Sec.~3 the condition by which the gauge invariant states are
identified, the so-called Gauss law constraint, is discussed. 
Sec.~4 presents the weak coupling limit of the transfer-matrix elements, together with
the subtleties related to the diverging group volume and its treatment. 
In Sec.~5 it is shown how the continuum part of the spectrum can be recovered 
in the small coupling limit of the lattice gauge model. Sec.~6 is devoted 
to concluding remarks. A detailed comparison between the lattice model in the weak coupling limit 
and the continuum model is presented in Appendix~A. 
The calculation of relevant eigenvalues is presented in Appendix~B.

\section{Review: Current Expansion in Fourier Basis}

In this section, the formulation of the transfer-matrix in the 
Fourier basis based on \cite{kiafat,vadfat} is shortly reviewed. 
In particular, the elements and notions emerged through the Fourier transform, specially in 
connection with the current interpretation of the Fourier modes, are emphasized.  
Following \cite{luscher,seiler}, the temporal gauge $A^0\equiv 0$ is used in the formulation.
The link `$\,l\,$' at site $\bm{r}$ in spatial direction `$\, i\,$' is represented by $l=(\bm{r},i)$. 
The gauge variables at adjacent times $n_t$ and $n_t+1$ are introduced 
to the model as follows:
\begin{align}\label{1}
n_t:& ~~~~\theta^{l}=a\,g\,A^{(\bm{r},i)}\\
\label{2}
n_{t}+1:& ~~~~\theta'^{l}=a\,g\,A'^{(\bm{r},i)}
\end{align}
taking values in $[-\pi,\pi]$ \cite{wilson,kogut}. 
Above, `$g$' and `$a$' are the gauge coupling and 
the lattice spacing parameters, respectively.
For a spatial lattice with $\nl$ number of links and
$\np$ number of plaquettes, it is helpful to define the 
plaquette-link matrix $\bm{M}$ of dimension $\np\times \nl$,
as following explicitly by its elements
\begin{align}\label{3}
	M^p_{~l}=\begin{cases}
		\pm 1,& \mbox{link $l=(\bm{r},\pm i)$ belongs to oriented plaquette $p$ } \\
		~~0,& \mbox{otherwise.}
	\end{cases}
\end{align}
\begin{figure}[t]
		\begin{center}
	\includegraphics[scale=2.2]{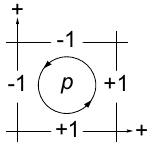}
		\end{center}
	\vskip -.5cm  
\caption{\small Graphical representation of definition (\ref{3}). }
	\label{fig1}
%	\vskip 1.4cm
\end{figure}
\noindent In Fig.~\ref{fig1} a graphical representation of the definition is given. Setting
\begin{align}\label{4}
	\gamma=\frac{1}{g^2}
\end{align}
the elements of the transfer-matrix $\widehat{V}$ are defined in terms of 
the Euclidean action between two adjacent times \cite{kiafat}
\begin{align}\label{5}
	\langle \bm{\theta'} |\widehat{V} | \bm{\theta}\rangle =~ \mathcal{A}~ &
	\prod_p \exp\!\left\{-\frac{\gamma}{2}\left[2-\cos\big(M^p_{~l}\,\theta^l\big)
	-\cos\big(M^p_{~l}\,\theta'^l\big)\right]\right\}\cr
	&\times\prod_{l}\exp\!\left\{-\gamma\Big[1-\cos\big(\theta^l-\theta'^l\big)\Big]\right\}
\end{align}
in which the summations over repeated indices are understood. 
In the above $\mathcal{A}$ is inserted to fix the normalization
\cite{normfixing}, an issue we will come back to it in Secs.~3 and 4.
The Fourier basis $| k_l \rangle$ is related to the compact $\theta$-basis by:
\begin{align}\label{6}
	\langle\theta^{l'}|k_{l} \rangle&=\frac{\delta^{l'}_{~l}}{\sqrt{2\pi}}
	\,\exp(\ri\,k_l\,\theta^l)	,~~~~~ 	k_l=0,\pm1,\pm2,\cdots
\end{align}
by which the transfer-matrix elements in the Fourier basis can be found
\begin{align}\label{7}
\langle\bm{k'}|\widehat{V}|\bm{k}\rangle
 = \frac{1}{(2\pi)^{\nl}}
\int_{-\pi}^\pi \prod_l  \rd\theta'_l \rd\theta_l \, e^{-\ri\,\bm{k'}\cdot\bm{\theta'}}
e^{\ri\,\bm{k}\cdot\bm{\theta}}\,\langle \bm{\theta'} |\widehat{V} | \bm{\theta}\rangle
\end{align}
Accordingly, it is shown that $\widehat{V}$ is block-diagonal in the Fourier basis 
\cite{vadfat,kiafat}, and all elements of a block can be presented by an arbitrary block's member 
$\bm{k}_\ast$ as the representative, whose co-blocks are all constructed as 
\begin{align}\label{8}
\bm{k}_{\ast\bm{q}}=\bm{k}_\ast + \qmb  
\end{align}
in which $\bm{q}$ is a row-vector with $\np$ integers as components. 
It will be seen later, as a manifestation of the current conservation, two co-blocks
can differ at most in the circulating currents inside plaquettes, allowing to have a non-zero 
matrix-element. Consequently, the matrix element between two co-blocks represented 
by $\bm{k}_\ast$ is found to be \cite{kiafat}
\begin{align}\label{9}
\langle\bm{k'}_{\ast\bm{q'}}|\widehat{V}|\bm{k}_{\ast\bm{q}}\rangle=
\mathcal{A}\,
 e^{-\gamma(\np+\nl) }  (2\pi)^{\nl}
\sum_{\{n^0_p\}}   \sum_{\{n_p\}}
& \prod_{p}\I_{q_p-n_p}\!\left(\frac{\gamma}{2}\right)\I_{q'_p-n_p+n^0_p}
\!\left(\frac{\gamma}{2}\right) 
\cr &
\prod_{l} \I_{k_\ast+\sum_p \!\! n_p M^p_{~l}}\!(\gamma)
\end{align}
in which $\bm{k}_{\ast\bm{q}}$ is given by (\ref{8}), and 
\begin{align}\label{10}
\bm{k'}_{\ast\bm{q'}}=\bm{k}_\ast + \qmpb 
\end{align}
In (\ref{9}), all summations are on integers, and $\I_r$'s are modified Bessel functions. Further, in a vector notation, 
$n^0_p$'s satisfy the following relation (including $\bm{n^0}=\bm{0}$) \cite{kiafat}:
\begin{align}\label{11}
\bm{n}^{\bm{0}}\cdot \bm{M}=\bm{0}.
\end{align}

Before proceeding to the strong coupling expansion, it is quite insightful to discuss
the physical meaning of the objects that emerged in the Fourier basis. 
First is the Fourier vector $\bm{k}$ itself, which can be directly understood by its appearance, 
namely by (\ref{7}), and its similar expression in the continuum theory, by 
the coupling of the current $J$ to the gauge field $A$ 
\begin{align}\label{12}
e^{\mathrm{i} \sum_l k_l \theta^l} = e^{\mathrm{i}\, a\,g\sum_l k_l A^l} \to
e^{\mathrm{i}\,g \int J\cdot A\, dx}
\end{align}
Above, $k_l$ is interpreted as the number of the current-quanta coupled to the 
gauge field $A^l$ associated to the link `$\,l\,$'. 
Accordingly, the current-vector $\bm{k}$ consists of the link-currents $k_l$'s. 
The integer value of $k_l$ reflects the fact that, thanks to the compact nature of gauge fields in the 
lattice model, the quantization of charge is satisfied automatically. 
By the above interpretation of $k_l$'s, the Fourier basis $|\bm{k}\rangle$ is representing 
the states of current quanta on the lattice links. 
Now, by the definition of the transfer-matrix $\widehat{V}=\exp(-a\widehat{H})$, with 
$\widehat{H}$ as the Hamiltonian, the matrix element $\langle\bm{k}'|\widehat{V}|\bm{k}\rangle$
is the transition amplitude between the states with $\bm{k}$ and $\bm{k'}$ currents 
during the imaginary time interval `$a$'. 

\begin{figure}[H]
	\begin{center}
		\includegraphics[scale=.6]{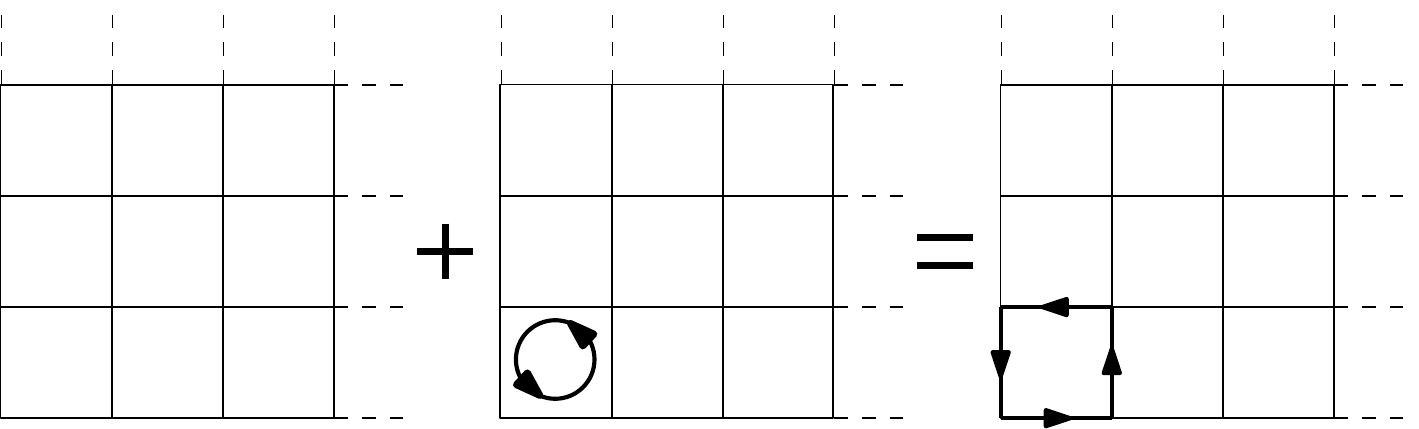}
%\vskip -.6cm   
\caption{\small The graphical representation of (\ref{13}) to construct $\bm{k_{0;1}}$ as 
co-block of $\bm{k_0}=\bm{0}$.}
\label{fig2}
\end{center}
\end{figure}

\begin{figure}[H]
	\begin{center}
		\includegraphics[scale=.6]{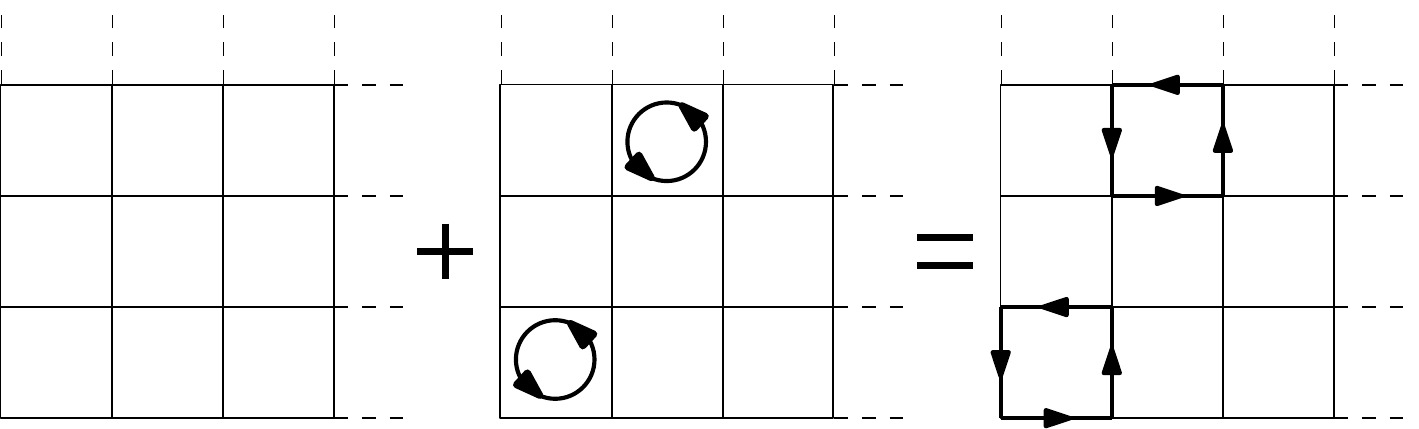}
%\vskip -.6cm   
\caption{\small Construction of a co-block of $\bm{k_0}$ with two non-adjacent plaquette-currents.}
\label{fig3}
	\end{center}
\end{figure}

To find the meaning of the integers $q_p$'s, first we need to realize the role of the matrix
$\bm{M}$, defined by its elements in (\ref{3}). In \cite{vadfat,kiafat}, an explicit representation 
of this matrix for the 2d spatial lattice is presented; see also the Appendices of the present work. 
Accordingly, by the definition (\ref{3}) each $q_p$
turns on two $+q_p$'s and two $-q_p$'s units of currents in four links of plaquette `$p$', 
being added to already existing currents of $\bm{k}_\ast$ in (\ref{8}) \cite{vadfat,kiafat}.
As an example, let us consider the block represented by the vacuum state $\bm{k}_\ast=\bm{k_0}=\bm{0}$,
and its co-block 
\begin{align}\label{13}
\bm{k_{0;1}}=\bm{k_0}+\bm{q_1}\cdot \bm{M}
\end{align}
with $\np$-component vector $\bm{q_1}=(1,0,\cdots,0)$. 
Eq.~(\ref{13}) is pictorially presented in Fig.~\ref{fig2}, showing that the resulting vector-current 
has only four links with non-zero current, namely two $+1$'s and two $-1$'s, making a 
circulating unit current in the first plaquette. 
 \vskip -.6cm
As seen, $q_p$ is determining the number of current quanta
circulating in the plaquette `$\,p\,$'. It is befitted to call the $q_p$ numbers as 
plaquette-currents or loop-currents.
As another example from the vacuum block, consider the state constructed by all
$q_p$'s zero, except two of non-adjacent plaquettes, as depicted in Fig.~\ref{fig3}. 
By above interpretation of $q_p$'s, the transfer-matrix elements are non-zero only between 
the current-states that differ in circulating currents inside plaquettes, as a consequence of the 
current conservation \cite{kiafat}.
 \vskip -.6cm

\begin{figure}[H]%[!ht]
	\begin{center}
		\includegraphics[scale=.6]{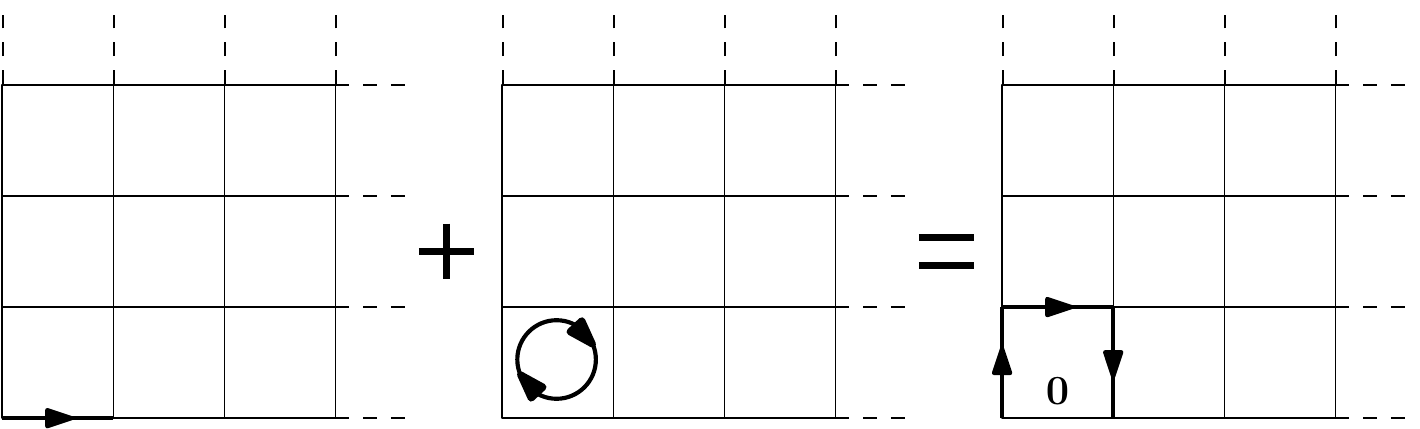}
%\vskip -.6cm   
\caption{\small $\bm{k_{1;-1}}=\bm{k_1}-\bm{q_1}\cdot \bm{M}$
as a co-block of $\bm{k_1}$ with three links having unit current.}
\label{fig4}
	\end{center}
\end{figure}

\begin{figure}[H]%[!ht]
	\begin{center}
		\includegraphics[scale=.6]{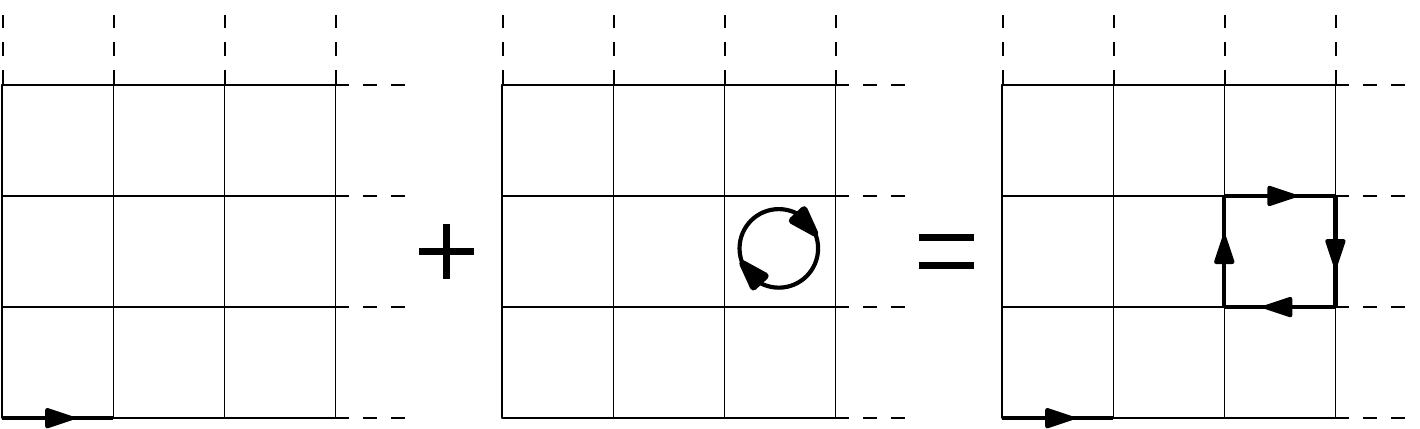}
%\vskip -.6cm   
\caption{\small A co-block of $\bm{k_1}$ with five links having unit current.}
\label{fig5}
	\end{center}
\end{figure}

Let us go beyond the vacuum block by considering the state represented by the 
$\nl$-component vector-current $\bm{k_1}=(1,0,\cdots,0)$,
which has one unit of current on the first link of the lattice, with all other link-currents zero. 
It is easy to see that there is no set of loop-currents $\bm{q}\cdot\bm{M}$ 
that could yield this vector from any member of the vacuum block. 
Two co-blocks of this state are presented in Figs. \ref{fig4} and \ref{fig5} \cite{kiafat}.
 \vskip -.5cm

 \vskip -.5cm

One of the most special blocks is the one represented by the 
vacuum state $\bm{k_0}=\bm{0}$. 
In \cite{vadfat} it is shown that, provided that the ground-state is unique,
it belongs to $\bm{k_0}$'s block.
The reason simply backs to the fact that in the extreme large coupling 
limit $g \to \infty$ ($\gamma \to 0$), as all elements except $V_{00}$ are
approaching zero, using the fact the energy and $\widehat{V}$ eigenvalues are related as 
\begin{align}\label{14}
E_i=-\frac{1}{a}\ln v_i 
\end{align}
the ground-state belongs to the vacuum block. By uniqueness of the 
ground-state, upon lowering the coupling, no crossing between ground-state 
by other states occurs, leaving the ground-states in the vacuum block at any coupling \cite{vadfat}.

The presented interpretation of $\bm{k}$ and $\bm{q}$ as link and loop currents 
leads to a diagrammatic strong coupling expansion of the transfer-matrix in the Fourier 
basis \cite{kiafat}. Specifically, it is shown that the elements of the transfer-matrix 
between two states of a block can be represented as a sum over occurrences of the virtual loop 
and link currents that transform both states to the vacuum \cite{kiafat}. 
In other words, the transition between two states, 
represented by the element $\langle \bm{k'} |\widehat{V} |\bm{k}\rangle$,
occurs as if the vacuum state is being passed as an intermediate state \cite{kiafat}.
The transformation to the vacuum via occurrences of the virtual currents is to be considered even for 
states that do not belong to the vacuum block. This is because that, the co-blocks
of a state are determined by adding loop-currents via $\bm{q}\cdot\bm{M}$ as (\ref{8}),
but the mentioned transform to the vacuum is due to both link and loop currents, the former via
`$\cos(\theta-\theta')$' term that is irrelevant for making co-blocks. It is due to these
link-currents that transformation of any state, including those from other blocks, into the vacuum 
is made possible \cite{kiafat}. The distinguished role of the vacuum 
state simply comes back to the fact that the Fourier integrals over $\bm{\theta}$ and $\bm{\theta'}$ 
related to both states are to be satisfied separately, namely $\bm{k}\to\bm{0}$ and 
$\bm{k'}\to\bm{0}$ \cite{kiafat}. 

To present the diagrammatic expansion, let us use the normalized transfer-matrix as:
\begin{align}\label{15}
\widehat{V}=\mathcal{A}\,  e^{-\gamma(\nl+\np) }  (2\pi)^{\nl}~\vred
\end{align}
by which $\langle\bm{0}|\vred|\bm{0}\rangle=1+\mathrm{O}(\gamma^2)$.
For two given states of $|\bm{k}\rangle$ and $|\bm{k'}\rangle$ in a block, consider
the case that they transform to the vacuum by $m$ and $m'$ numbers of the virtual loop-currents, 
respectively, accompanied by $\ell$ numbers of the virtual link-currents for both states. 
Now, the numerical factor associated 
with the matrix-element of transition through the mentioned transform is \cite{kiafat}
\begin{align}\label{16}
\left[\langle\bm{k'}|\vred|\bm{k}\rangle\right]_{m,m',\ell} =
 \mathcal{K}_{m,m',\ell}~  \frac{1}{2^{2m+2m'+\ell}}\frac{1}{m!\,m'!\,\ell!} \,\gamma^{m+m'+\ell}
\end{align}
in which $\mathcal{K}_{m,m',\ell}$ is the combinatorial factor representing the number of ways that
loop and link currents can be combined, regarding the initial and final transformations to the vacuum.
As mentioned, the above matrix-elements can be represented by a set of 
graphs in which a proper combination of \textit{virtual} loop and link currents would make the required 
pass through the vacuum. The ways that the initial and final states transform
to the vacuum, accompanied by the associated numerical and combinatorial factors of 
each transition, simply fix the terms in the strong coupling expansion \cite{kiafat}.
In this respect, these graphical representations 
can serve as the Feynman diagrams in the perturbative quantum field theory. 
The elements being used in graphs are simply the currents, loop or link ones, being 
characterized by their real or virtual natures. Accordingly, the initial and final states, being 
determined only by the link-currents, are interpreted as real and presented by a combination 
of solid lines as 
\begin{align} \label{17}
\linkcur~~~~~~\mathrm{or}~~~~~~~\linkcurinv
\end{align}
and their rotated versions. 
Instead, the loop and link-currents occurred during the transforms
are interpreted as virtual, being drawn in dashed form as below for the loop currents
\begin{align}\label{18}
\circ ~~~~~~~~\mathrm{or}~~~~~~~~ \circinv
\end{align}
and the following for link-currents
\begin{align}\label{19}
\linkdash~~~~~~~\mathrm{or}~~~~~~~\linkdashinv
\end{align}
To emphasize the pass through the vacuum, we use the notation 
$\bm{k'}\displaystyle{\mathop{\to}^0}\,\bm{k}$ for the matrix-element 
$\langle\bm{k'}|\vred|\bm{k}\rangle$ \cite{kiafat}. As examples of the 
diagrammatic representation of the terms in the strong coupling expansion, 
the diagrams contributing to the vacuum to vacuum  (v.t.v.) transition at 
order $\gamma^2$  are
\begin{align}\label{20}
\Big[\langle \bm{0}|\vred|\bm{0}\rangle_{\bm{0}}&\Big]_{\gamma^2}:
\begin{cases}
\bm{0}+\twocirc ~ \pmb{\longrightarrow} \bm{0}&~ \frac{1}{2^4}\frac{1}{2!} 2\np\cr
\bm{0} ~ \pmb{\longrightarrow} \bm{0}+\twocirc&~ \frac{1}{2^4}\frac{1}{2!} 2\np\cr
\bm{0}+ \backforth~ \pmb{\longrightarrow} \bm{0}+\backforth &~ \frac{1}{2^2}\frac{1}{2!} 2\nl
\end{cases}
\end{align}
in which as before, $\np$ and $\nl$ are number of plaquettes and link in the lattice, respectively.
At order $\gamma^4$, denoting
\begin{align}\label{21}
C^m_n=\left(\begin{matrix}n \cr m \end{matrix}\right) = \frac{n!}{m!(n-m)!}
\end{align}
we have the following
\begin{align}\label{22}
\Big[\langle \bm{0}|\vred|\bm{0}\rangle_{\bm{0}}\Big]_{\gamma^4}:
\begin{cases}
\bm{0}+\twocirc~\twocirc ~\displaystyle{\mathop{\pmb{\longrightarrow}}^0}~ \bm{0} &~ \frac{1}{2^8}\frac{1}{4!}C^2_4 (2\np^2-\np) \cr
\bm{0} ~\displaystyle{\mathop{\pmb{\longrightarrow}}^0}~ \bm{0}+\twocirc~\twocirc &~ \frac{1}{2^8}\frac{1}{4!}C^2_4 (2\np^2-\np) \cr
\bm{0} +\twocirc~\displaystyle{\mathop{\pmb{\longrightarrow}}^0}~ \bm{0}+\twocirc &~ \frac{1}{2^8}\frac{1}{2!2!}C^1_2C^1_2 \np^2 \cr
\bm{0} +\twocirc~\backforth ~\displaystyle{\mathop{\pmb{\longrightarrow}}^0}~ \bm{0}+\backforth &~ \frac{1}{2^6}\frac{1}{2!2!}C^1_2C^1_2 \np\nl \cr
\bm{0} +\backforth ~\displaystyle{\mathop{\pmb{\longrightarrow}}^0}~ \bm{0}+\twocirc~\backforth &~  \frac{1}{2^6}\frac{1}{2!2!}C^1_2C^1_2 \np\nl \cr
\bm{0} +\backforth~\backforth ~\displaystyle{\mathop{\pmb{\longrightarrow}}^0}~ \bm{0}+\backforth~\backforth &~ \frac{1}{2^4}\frac{1}{4!}C^2_4 (2\nl^2-\nl) 
\end{cases}
\end{align}
All together we have \cite{kiafat}
\begin{align}\label{23}
\langle \bm{0}|\vred|\bm{0}\rangle_{\bm{0}}=& 1+ \left(\frac{\np}{8} + \frac{\nl}{4}\right) \gamma^2 \cr
&+\left(-\frac{\nl}{64}+\frac{\nl^2}{32}-\frac{\np}{512}+ \frac{\np \nl}{32} +\frac{\np^2}{128}
\right)\gamma^4 + \mathrm{O}(\gamma^6)
\end{align}
Similarly, denoting $\bm{k_{0;1}}$ as $|\bm{1}\rangle$, 
for the transition $\bm{1}\displaystyle{\mathop{\to}^0}\,\bm{1}$ we have 
\begin{align}\label{24}
\Big[\langle \bm{1}|\vred|\bm{1}\rangle_{\bm{0}~}\Big]_{\gamma^2}:
\recdashcirc  \mathop{\pmb{\longrightarrow}}^0 \recdashcirc  ~~~~~~~\frac{\gamma^2}{16}
\end{align}
and 
\begin{align}\label{25}
\left[\langle \bm{1}|\vred|\bm{1}\rangle_{\bm{0}}\right]_{\gamma^4}:
\begin{cases}
 \recdashcirc  ~\displaystyle{\mathop{\pmb{\longrightarrow}}^0}~ \recdashcirc ~ \twocirc &~ \frac{1}{2^8}\frac{1}{3!} C^1_3 (2\np-1) \cr
\recdashcirc ~ \twocirc ~\displaystyle{\mathop{\pmb{\longrightarrow}}^0}~ \recdashcirc    &~ \frac{1}{2^8}\frac{1}{3!}C^1_3 (2\np-1) \cr
\recdashcirc~\backforth ~\displaystyle{\mathop{\pmb{\longrightarrow}}^0}~ \recdashcirc  ~ \backforth &~ \frac{1}{2^6}\frac{1}{2! } C^1_2 \nl \cr
\recdashrec ~\displaystyle{\mathop{\pmb{\longrightarrow}}^0}~ \recdashrec  &~ \frac{1}{2^4}\frac{1}{4! } C^2_4 \,4
\end{cases}
\end{align}
leading to 
\begin{align}\label{26}
\langle \bm{1}|\vred|\bm{1}\rangle_{\bm{0}}= \frac{\gamma^2}{16}
+\left(\frac{15}{256}+\frac{\nl}{64}+\frac{\np}{128}\right)\gamma^4 + \cdots 
\end{align}
As final sample of the strong coupling expansion we find for the 
$\bm{1}\displaystyle{\mathop{\to}^0}\,\bm{0}$ transition
\begin{align}\label{27}
\langle \bm{1}|\vred|\bm{0}\rangle_{\bm{0}}=&\frac{\gamma}{4}
+\left(\frac{-1}{128}+\frac{\np}{32}+\frac{\nl}{16}\right)\gamma^3\cr
&+\left( \frac{49}{3072}-\frac{3\nl}{512}+\frac{\nl^2}{128}-
\frac{3\np}{2048}+\frac{\nl\np}{128}+\frac{\np^2}{512} \right) \gamma^5+\cdots
\end{align}
Many other examples of the diagrammatic expansion in the strong coupling limit are presented in \cite{kiafat}. 
For a general matrix element, it can be shown that two subsequent orders of $\gamma$ 
in the expansion of an element differ by 2. 
This simply comes back to the fact that, any given order of an element differs 
from a higher order one by adding an even number of the virtual currents, as required by the transforms
of states to the vacuum \cite{kiafat}. 
So the expansion for an element in $\bm{k}_\ast$-block looks like 
\begin{align}\label{28}
\langle\bm{k'}_{\ast\bm{q'}}|\vred|\bm{k}_{\ast\bm{q}}\rangle_{\bm{k}_\ast} = \gamma^h \,
	(c_0 + c_{2}\,\gamma^{2}+ c_{4}\,\gamma^{4} +c_{6}\,\gamma^{6}+\cdots)
\end{align}
in which `$h$' is the lowest order at which the transforms of both initial and final states 
to the vacuum are made possible. As the consequence, the expansion is in powers of $1/g^4$, 
which makes it fairly reliable for even not so large `$g$'. The value of 
`$h$' can be determined as well, once $\bm{k}_\ast$, $\bm{q}$ and 
$\bm{q'}$ are given. To make things systematically, we use the convention that
the representative vector would have the minimum value of 
\begin{align}\label{29}
|\bm{k}_\ast|=\sum_{l=1}^{\nl} |k_{\ast l}|
\end{align}
The value of `$h$' is determined once 
$\bm{q}$ and $\bm{q'}$ of (\ref{8}) and (\ref{10}) are given, as follows \cite{kiafat}
\begin{align}\label{30}
h=|\bm{k}_\ast|+|\bm{q}|+|\bm{q'}|
\end{align}
in which 
\begin{align}\label{31}
|\bm{q}|=\sum_{p=1}^{\np} |q_p|,~~~~~|\bm{q'}|=\sum_{p=1}^{\np} |q'_p|
\end{align}

\section{Gauss Law Constraint in Fourier Basis}
It is known that in the procedure of quantization of lattice gauge theories,
due to the compact nature of gauge fields, the gauge fixing condition
is not needed in the path-integral \cite{wilson,kogut}. However, it is still
necessary to restrict the available states to the gauge-invariant ones, 
represented by the so-called Wilson loops in the field basis\cite{wilson,kogut}. 
The same is true in the temporal gauge used to derive the 
transfer-matrix elements in the Fourier basis \cite{vadfat,kiafat}.
That is so because, as the result of partial gauge fixing condition 
$A^0\equiv 0$, the model still enjoys residual gauge symmetry,
acting as time-independent (spatial) gauge transformations \cite{cruetz,smit}. 
In the case of the temporal gauge, it is known that the gauge invariance condition 
on the states is nothing but the Gauss law constraint \cite{smit,cruetz}. 
In other words, the generator of the spatial gauge transformation, appearing in
the Gauss law, should leave the physical states unaffected \cite{smit,cruetz}. 
Accordingly, it is shown that in the temporal gauge the set of physical states consists of 
the Wilson loops entirely lying in the spatial directions \cite{smit}. 
In the field basis, the wave-function corresponding to the spatial Wilson loop 
is represented by the path-ordered exponential around the spatial closed loop `C' as \cite{wilson,smit}
\begin{align}\label{32}
\psi_\mathrm{C}[\bm{\theta}]= \exp\!\Big[\mathrm{i}
\sum_{l \,\in\, \mathrm{C}} \! J_{l}\, \theta^l \Big]
\end{align}
in which, depending on the direction of `C' on link $l$, $J_l=\pm J$ \cite{wilson}.
It is obvious that by construction (\ref{32}) is gauge-invariant. The most general gauge-invariant 
wave-functions are simply constructed by multiplications of those like (\ref{32}) for a single loop. 
It is the purpose of this section to map 
the above physical states in the field basis to the Fourier basis. The transform to 
the Fourier basis is simply 
\begin{align}\label{33}
\widetilde\psi_\mathrm{C}[\bm{k}]=\frac{1}{(2\pi)^{\nl/2}}
\int_{-\pi}^\pi  
\prod_l  \rd\theta_l \, \exp\!\Big[\mathrm{i}\Big(
\sum_{l\, \in\, \mathrm{C}}\! J_{l}\,\theta^{l} - \bm{k}\cdot\bm{\theta}\Big) \Big]
\end{align}
leading to the following as the product of Kronecker $\delta$'s of integer values:
\begin{align}\label{34}
\widetilde\psi_\mathrm{C}[\bm{k}]=(2\pi)^{\nl/2} \prod_{l \,\in\, \mathrm{C}} \!\delta(k_l-J_{l}) \,
\prod_{l'\notin \mathrm{C}} \delta(k_{l'})
\end{align}
by which the gauge invariant state in the Fourier basis has only non-zero currents $|J_l|=J$ 
along the closed spatial loop `C' of $\psi_\mathrm{C}[\bm{\theta}]$. 
The identification of the above states is particularly simple by the notions presented in the previous section. 
For the case of the 1d periodic spatial lattice, the only closed loop is the one around the entire the 
periodic spatial direction. As in this case, the transfer-matrix is diagonal \cite{itzyk,wipf}, and
the element for a periodic lattice with $\nl$ links finds the form:
\begin{align}\label{35}
\langle k|\widehat{V}|k\rangle= \mathcal{A}\, \big(e^{-\gamma}\, \I_k(\gamma)\big)^{\nl},
~~~~~k=0,\pm1,\pm2,\cdots
\end{align}
where $\mathcal{A}$ is inserted to fix the normalization \cite{normfixing}.
By the elements of diagonal $\widehat{V}$, the exact spectrum of 1d model is found \cite{itzyk, wipf}:
\begin{align}\label{36}
E_k=-\frac{1}{a}\ln \langle k|\widehat{V}|k\rangle 
\end{align}
Above, due to the property of the Bessel functions $\I_0>\I_{\pm 1}>\I_{\pm 2}\cdots$,
the ground-state is given by $k=0$ \cite{itzyk, wipf}. 
In \cite{normfixing}, the normalization $\mathcal{A}$ is fixed in a such way that in the weak coupling
limit $\gamma \gg 1$, using the asymptotic behavior of Bessel functions: 
\begin{align}\label{37}
\I_k(\gamma)\simeq \frac{e^\gamma}{\sqrt{2\pi \gamma}}\,
e^{-(k^2-1/4)/(2\gamma)}\,\left[1+\mathrm{O}\left(\frac{1}{\gamma^2}\right)\right]
,~~~~~~~~~~~\gamma\gg k
\end{align}
the continuum energy density (energy per link) expected by the classical model is recovered
\begin{align}\label{38}
\frac{E_k}{\nl} \simeq \frac{(k^2-\sfrac{1}{4}) g^2}{2\,a},~~~~~ g\ll 1,~~k\in \mathbb{Z}.
\end{align}
For the case of an infinite 1d lattice, one still can define $\widetilde\psi_\mathrm{C}[k]$ along
the spatial direction, by which the above finite continuum energy density is valid in 
the limit $\nl\to\infty$.

For the case of more than one dimension, it is still possible to consider the current-loops 
around the entire periodic directions, such as in Figs.~6 and 7. 
As will be seen in the next section, the blocks owners of these periodic large current-loops
are again responsible for the continuum spectrum expected in the weak coupling limit.
However, in the strong coupling limit $\gamma\ll 1$, these periodic loop-currents and 
their blocks contribute only to the extremely excited energies \cite{kiafat}. This can be understood 
easily by the strong coupling expansion of \cite{kiafat}, reviewed in the previous section. 
Accordingly, the matrix element associated with the mentioned states behaves as $\gamma^{\nl}$ or 
with higher powers in the limit $\gamma\ll1$, leading to extremely high energies in the large size limit 
$\nl\gg1$ by $E=-\nl \ln \gamma$.

\begin{figure}[H]%[!ht]
	\begin{center}
		\includegraphics[scale=.8]{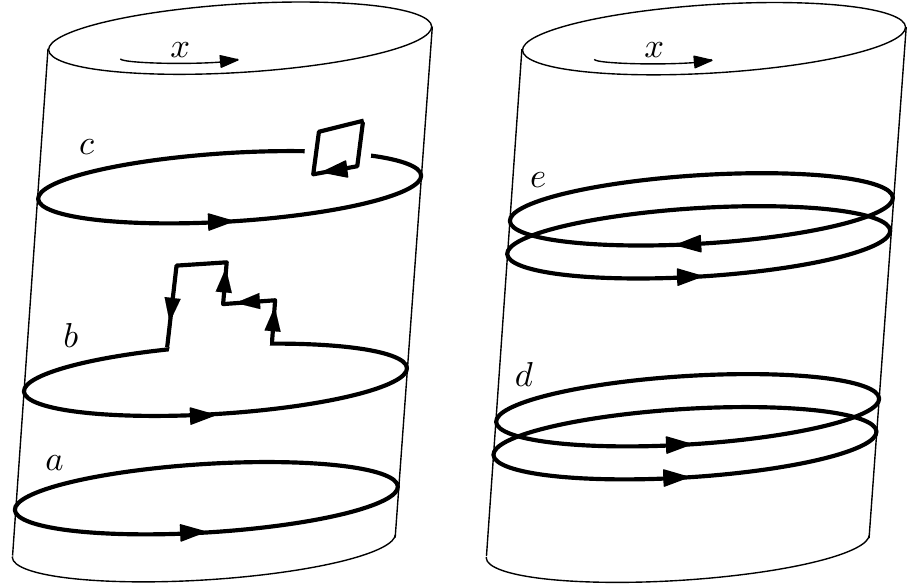}
%\vskip -.6cm   
\caption{\small Left: the periodic loop-current $a$, and two of its co-blocks $b$ and $c$.
Right: $d$ as a two spaced equal periodic loop-currents, and $e$ as a two spaced opposite periodic 
loop-currents. In fact, $e$ belongs to the vacuum block.}
\label{fig6}
	\end{center}
\end{figure} \vskip -.5cm

\begin{figure}[H]%[!ht]
	\begin{center}
		\includegraphics[scale=.7]{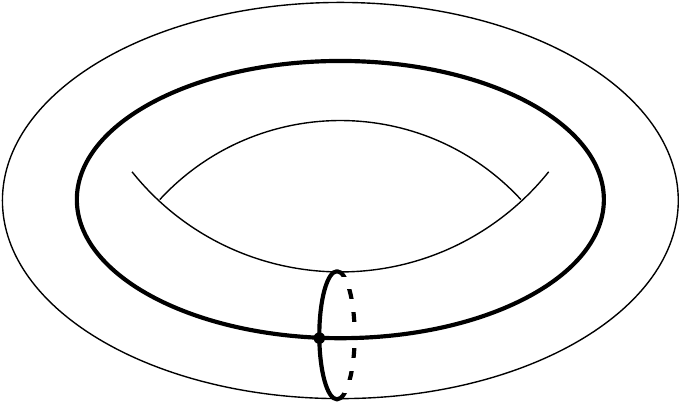}
%\vskip -.6cm   
\caption{\small A gauge invariant Fourier state that consists of two periodic loop-currents 
along two spatial directions. The two currents are not necessarily equal.}
\label{fig7}
	\end{center}
\end{figure} \vskip -.5cm

Apart from above mentioned loops along periodic directions, there are other loops with  
less number of links in more than one dimension as well. Examples of such closed currents in the 2d 
spatial lattice are given in Fig.~8. 
Further examples in a 3d lattice, such as currents at edges are given in Fig.~9, and in general form in Fig.~10.
All of these closed currents can be constructed by superposition of equal plaquette-currents of the previous 
section, as shown in each case. 
As a consequence, all of these gauge invariant finite size closed currents belong to the vacuum block.
In the previous section, it is seen how the above closed-currents are generated by 
adding the plaquette-currents to the vacuum. 
By the strong coupling expansion of the previous section, a closed current 
generated by $n_c$ number of the plaquette-currents leads to the matrix element
$\langle n_c |\vred|n_c\rangle \simeq \gamma^{2n_c}$ in the limit 
$\gamma\ll 1$. This leads to the fact that, in the strong coupling regime, the 
spectrum of the model comes effectively from the vacuum block; as mentioned before, 
the other blocks of the periodic loop-currents generate only extremely excited energies.
As shown in \cite{kiafat}, by means of simple perturbative methods,
the eigenvalues of the vacuum block can be evaluated in the strong coupling regime \cite{kiafat}.

\begin{figure}[H]%[!ht]
	\begin{center}
		\includegraphics[scale=.5]{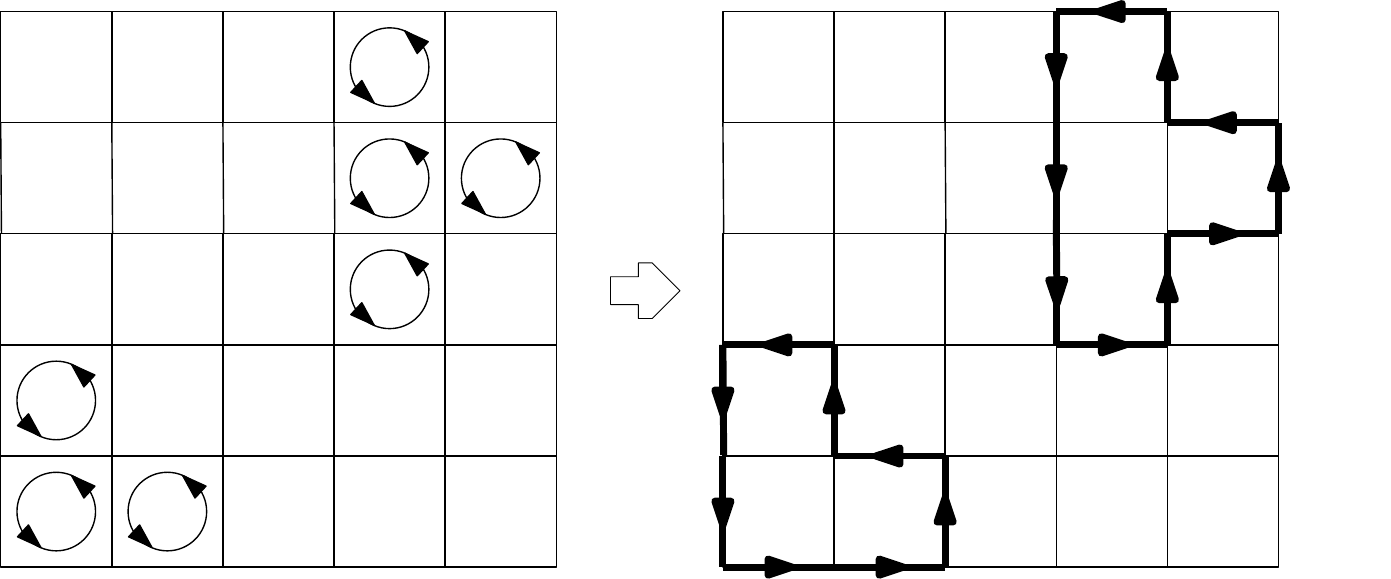}
%\vskip -.6cm   
\caption{\small Examples of 2d gauge invariant closed currents from the vacuum block. }
\label{fig8}
	\end{center}
\end{figure} \vskip -.5cm

\begin{figure}[H]%[!ht]
	\begin{center}
		\includegraphics[scale=.75]{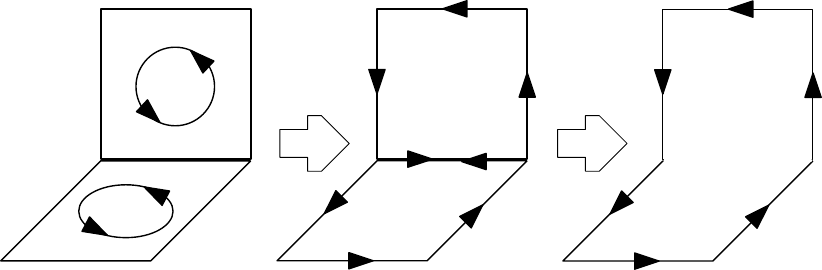}
\vskip .5cm
		\includegraphics[scale=.7]{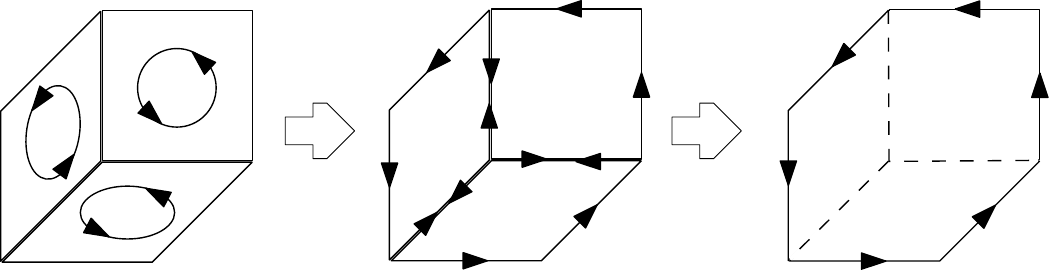}
%\vskip -.6cm   
\caption{\small Examples of 3d closed edge currents from the vacuum block. }
\label{fig9}
	\end{center}
\end{figure} \vskip -.5cm

\begin{figure}[H]%[!ht]
	\begin{center}
		\includegraphics[scale=.6]{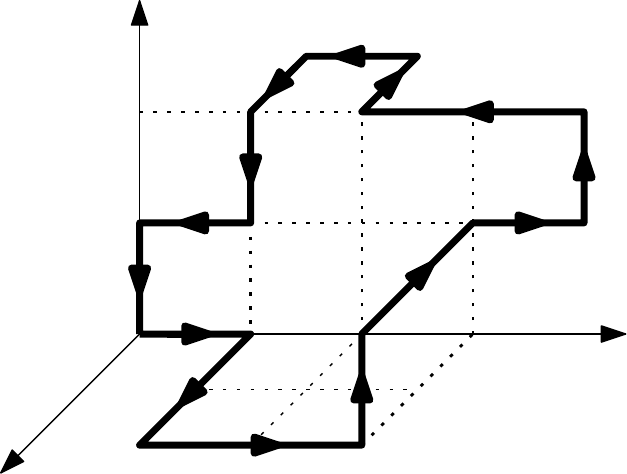}
\hskip 1cm 
		\includegraphics[scale=.6]{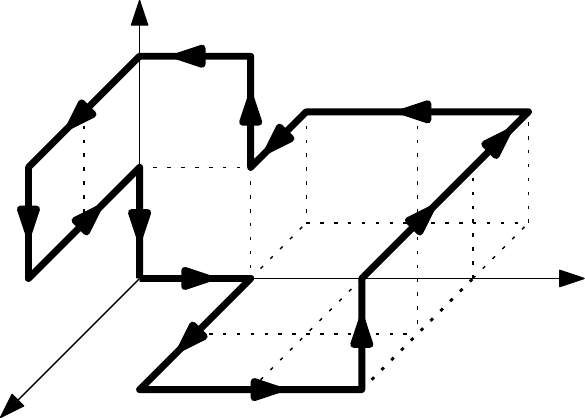}
%\vskip -.6cm   
\caption{\small General 3d closed currents from the vacuum block.}
\label{fig10}
	\end{center}
\end{figure} \vskip -.5cm

\section{Transfer-Matrix in Weak Coupling Limit}
The quantum theory of pure U(1) gauge model in the presence of source $J_\mu$ may be defined by 
means of the path-integral 
\begin{align}\label{39}
\mathcal{Z}[J]=\int \mathscr{D}\! A\, \exp\!\left[\mathrm{i}
\int \! \rd x \left(-\frac{1}{2}A^\mu
(\eta_{\mu\nu} \Box - \partial_\mu \partial_\nu)  A^\nu
 + J_\mu A^\mu  \right) \right]
\end{align}
leading to
\begin{align}\label{40}
\mathcal{Z}[J]\propto \left|\det(\eta_{\mu\nu} \Box - \partial_\mu \partial_\nu)\right|^{-1/2}
\end{align}
However, the above expression is infinite due to the zero modes of the operator, corresponding to the 
pure gauge configurations $A_\mu=\partial_\mu f(x)$:
\begin{align}\label{41}
(\eta_{\mu\nu} \Box - \partial_\mu \partial_\nu)\partial^\nu f(x) = \Box\, \partial_\mu f 
- \partial_\mu\, \Box f  =0
\end{align}
The above mentioned infinity, in other words, is due to the infinite group volume, resulting from 
integration over $[-\infty,\infty]$ of redundant gauge degrees.
Accordingly, the common recipe to avoid the divergent behavior is to fix the gauge. 
Among many possibilities, one may impose the temporal gauge used in the present work 
$A^0= 0$, which is known as an incomplete gauge due to the residual gauge freedom.  
To make the gauge fixing complete, the so-called Coulomb gauge condition by 
$\bm{\nabla}\cdot \bm{A}= 0$ may be added. 

It is commonly expressed that in the lattice formulation of gauge theories, the above 
mentioned gauge fixing is not necessary. This simply is due to the compact nature 
of gauge fields  $-\pi/g \leq a\,A\leq \pi/g$, by which the integration over unfixed gauge degrees would not lead 
to infinity. In going from the lattice theory to continuum, however, one has to care
about the diverging integrals. The main purpose of the present section is to show
in detail how the infinities emerge in going to the extreme weak coupling limit 
$g\to 0$ of the lattice model. 
In particular, it is seen how the group volume emerges through integration over the zero modes 
of the operator by the quadratic action in $g\to 0$ limit.
The number of mentioned zero modes is equal to the number of ``fixing" spots on lattice by the
condition $\bm{\nabla}\cdot \bm{A}=0$, and also the number of pure gauge field configurations,
both equal to the number of the lattice sites. 

In the weak coupling limit $g\ll 1$ the configurations with $\theta \ll 1$ find 
the dominant contribution. So the cosines in the action can be expanded, leading to the 
quadratic form similar to the continuum model of gauge theory. Back from the angle variable 
$\theta=a\,g A$ to the original variables `$A$', inserted in the following vector with $2\nl$ components 
\begin{align}\label{42}
\bm{\eta}= \begin{pmatrix}\bm{A} \cr \bm{A'} \end{pmatrix}
\end{align}
the elements of the transfer-matrix are given by means of the following quadratic action
\begin{align}\label{43}
	\langle \bm{\theta'} |\widehat{V} | \bm{\theta}\rangle = \mathcal{A}\, 
\exp\!\left(-\frac{a^2}{2}\,\bm{\eta}^T \bm{C}\, \bm{\eta} +\mathrm{O}(g^2)  \right)
\end{align}
in which matrix $\bm{C}$ has $2\nl\times 2\nl$ dimensions, given explicitly by 
\begin{align}\label{44}
\bm{C}=&\begin{pmatrix} 1 & -1 \cr 
-1 & 1  \end{pmatrix}
\otimes \mathbb{1}_\mathrm{L} + \frac{1}{2} \, \mathbb{1}_\mathrm{2} \otimes \bm{M}^T\bm{M}	
\\
\label{45}
=&	\left(\begin{array}{c|c}
	\mathbb{1}_\mathrm{L}+\frac{1}{2}\bm{M}^T\bm{M} ~ &  -\mathbb{1}_\mathrm{L} \cr 
	\hline 
   -\mathbb{1}_\mathrm{L}   &   	\mathbb{1}_\mathrm{L}+\frac{1}{2}\bm{M}^T\bm{M}
	\end{array}\right)
\end{align}
with $\mathbb{1}_2$ and $\mathbb{1}_\mathrm{L}$ as two and $\nl$ dimensional 
identity matrices, respectively.
From now on we set $a=1$, which can be recovered easily by dimensional considerations.
To transfer to the Fourier basis, we define the vector
\begin{align}\label{46}
\bm{\kappa}=\begin{pmatrix}\bm{k} \cr -\bm{k'} \end{pmatrix}
\end{align}
by which, using (\ref{6}) with original `$A$' variables, we have 
\begin{align}\label{47}
	\langle \bm{k'} |\widehat{V} | \bm{k}\rangle \simeq \, \mathcal{A}\,\frac{g^{2\nl}}{(2\pi)^{\nl}}\, 
\int^{\pi/g}_{-\pi/g}	\rd\bm{\eta}\,
\exp\!\left(-\frac{1}{2}\bm{\eta}^T \bm{C}\, \bm{\eta} + \mathrm{i}\,g\, \bm{\eta}^T \bm{\kappa}
\right)
\end{align}
In the limit $g \to 0$ the above integral is practically the Gaussian one. 
However, at first the zero eigenvalues of  matrix $\bm{C}$ should be treated.
The origin of these zero eigenvalues simply goes back to the fact that, in the 
present temporal gauge $A_0\equiv 0$, there are still unfixed gauge freedoms.
These unfixed degrees contribute infinitely due to the 
volume of the group in the uncompact limit $g \to 0$. 
The matrix $\bm{C}$ is symmetric, so there is a basis in which it is diagonal. 
In fact, by the eigenvectors of $\bm{C}$ one can construct the matrix $\bm{P}$,
by which
\begin{align}\label{48}
\bm{\tilde{C}}&=\bm {P}^{-1} \bm{C}\bm{P}\\
\label{49}
\bm{\tilde{\eta}}&=\bm{P}^{-1}\bm{\eta}\\
\label{50}
\bm{\tilde{\eta}}^T&=\bm{\eta}^T \bm{P}\\
\label{51}
\bm{\tilde{\kappa}}&=\bm{P}^{-1}\bm{\kappa}
\end{align}
with $\bm{\tilde{C}}$ being diagonal. 
By the expectations from the continuum model, we expect that
the matrix $\bm{C}$ has zero eigenvalues. These eigenvalues corresponds to
the pure gauge field configurations as (\ref{41}), by which $\bm{E}=\bm{B}=\bm{0}$.
Accordingly, the diagonal matrix may be represented in the following way
\begin{align}\label{52}
	\bm{\tilde{C}}=	\left(\begin{array}{c|c}
		~~   &   	~~~~ \cr
		~~~\bm{\tilde{C}}_\mathrm{u}~~~~ &  ~\bm{0}~ \cr 
				~~   &   	~~~~ \cr
		\hline 
       ~\raisebox{1mm}{$\bm{0}$}~   &  \raisebox{7mm}{$~$}\bm{\tilde{C}}_\mathrm{d} 
	\end{array}\right)
\end{align}
where $\bm{\tilde{C}}_\mathrm{u}$ is diagonal, and $\bm{\tilde{C}}_\mathrm{d} =\bm{0}$.
The dimension of the subspace by the zero modes determines the size of block
$\bm{\tilde{C}}_\mathrm{d} =\bm{0}$. 
As zero modes are being represented by the pure gauge configurations as 
$\bm{A}(\bm{x})=\bm{\nabla}f(\bm{x})$, the size is expected to be the number of possible
configurations in the coordinate $\bm{x}$-space, being effectively the number of sites in the 
lattice version of the model. 
For the periodic spatial lattices with $N_s$ sites in each 
direction, by the explicit representation of the matrix $\bm{M}$,
one can check that it is in fact the case, as summarized in Tab.~1. 
In Appendix~A a detailed comparison between the lattice model in the weak coupling limit 
and the continuum model is presented. In particular, using the 
explicit representations of matrices $\bm{C}$ and $\bm{M}$ for 2d and 3d models, 
it is shown how the models on the  lattice and continuum act similarly in the above mentioned respects. 

\vskip .2cm
\begin{table}[H]
  \begin{center}
    \label{tab:table1}
\begin{small}     \begin{tabular}{c|c|c|c}
 & No. of sites & No. of links & dim. of  $\bm{\tilde{C}}_\mathrm{d}$ \\
      \hline
 2d lattice & $N_s^2$ & $\nl=2N_s^2$ & $\nd=N_s^2+1$ \\ \hline
 3d lattice & $N_s^3$ & $\nl=3N_s^3$ & $\nd=N_s^3+2$ 
    \end{tabular}
\end{small}
\caption{\small For  periodic lattices in two and three dimensions the size of block $\bm{\tilde{C}}_\mathrm{d}$
is given by explicit representations of $\bm{M}$.}
  \end{center}
\end{table}
\vskip -.2cm

The projection of $\bm{\tilde{\kappa}}$ on the sub-space by $\bm{\tilde{C}}_\mathrm{d}$, 
denoted by $\bm{\tilde{\kappa}}_\mathrm{d}$, do not appear in the quadratic part  
$\bm{\eta}^T \bm{C}\, \bm{\eta}$. Back to the angle variables $\theta=g\,A$ for
these zero modes, upon integration over $\theta\in [-\pi,\pi]$, 
the Kronecker $\delta$'s are developed, leading to
\begin{align}\label{53}
\langle \bm{k'} |\widehat{V} | \bm{k}\rangle = \mathcal{A}\frac{g^{2\nl-\nd}}{(2\pi)^{\nl-\nd}} 
\,\delta(\bm{\tilde{\kappa}}_\mathrm{d})
\int^{\pi/g}_{-\pi/g} \rd\bm{\tilde{\eta}}_\mathrm{u}\,
\exp\!\left(-\frac{1}{2}\bm{\tilde{\eta}}_\mathrm{u}^T \bm{\tilde{C}}_\mathrm{u}\, \bm{\tilde{\eta}}_\mathrm{u} 
+ \mathrm{i}\,g\, \bm{\tilde{\eta}}_\mathrm{u}^T \bm{\tilde{\kappa}}_\mathrm{u}
\right)
\end{align}
in which only integrals on non-zero modes are left. 
By the explicit representation of $\bm{M}$, it is an easy task to see that 
$\delta(\bm{\tilde{\kappa}}_\mathrm{d})$ is automatically satisfied for members of 
a block of the transfer matrix, constructed by the relations (\ref{8}) and (\ref{10}). 
So the only remaining job is the integration over the non-zero modes, which 
is well approximated by Gaussian integrals in the limit $g\to 0$, 
using the following relation \cite{erf1,erf2}:
\begin{align}\label{54}
\mathrm{erf}(z) &= \frac{2}{\sqrt{\pi}} \int_0^z e^{-x^2}\rd x &   \cr
&= 1-\frac{e^{-z^2}}{\sqrt{\pi}} \sum_{n=0}^\infty 
\frac{(-1)^n (2n-1)!!}{2^n\, z^{2n+1}}   & z\gg 1
\end{align}
Accordingly, one has the following for $\Lambda \gg 1$ \cite{erf1}:
\begin{align}\label{55}
\int_{-\Lambda}^\Lambda e^{-\frac{1}{2} \vec{x}^T\! \bm{F}\, \vec{x} 
+\mathrm{i}\,
\vec{B}\cdot \vec{x} }\,\rd^nx = \sqrt{\frac{(2\pi)^n}{\det \bm{F}}}\, e^{-\frac{1}{2} \vec{B}^T \bm{F}^{-1} \vec{B}} 
+ \mathrm{O}\left(\frac{e^{-\Lambda^2/2}}{\Lambda}\right).
\end{align}
As mentioned earlier, to satisfy $\delta(\bm{\tilde{\kappa}}_\mathrm{d})$ two states must 
belong to the same block of the transfer-matrix. For the states (\ref{8}) and (\ref{10}) 
in the block represented by $\bm{k}_\ast$, we find the following explicit form
\begin{align}\label{56}
\langle\bm{k'}_{\ast\bm{q'}}|\widehat{V}|\bm{k}_{\ast\bm{q}}\rangle%_{\bm{k}_\ast}
= \mathcal{A}& \frac{g^{2\nl-\nd}}{(2\pi)^{\nl-\nd}} 
%\delta(\bm{\kappa}_\mathrm{down})
\sqrt{\frac{(2\pi)^{2\nl-\nd}}{\det \bm{\tilde{C}}_\mathrm{u}}} \cr
&\times \Bigg[ \exp\!\left(\!-\frac{g^2}{2}
\bm{\tilde{\kappa}}_\mathrm{u}^T \bm{\tilde{C}}_\mathrm{u}^{-1} \bm{\tilde{\kappa}}_\mathrm{u} \! \right) 
+  \mathrm{O}\!\left(g\,e^{-\pi^2/g^2}\right) \! \Bigg], ~~~~~~~g\ll 1
\end{align}
The resulted expression presents the exact dependence of matrix-elements on
the lattice size parameters $\nl$ and $\nd$, related to the number of sites $N_s$ as given in Tab.~1.
This is due to the specific parametrization, by which the derivations and different limits are made possible 
for any lattice size. In the next section an example of large lattice limit for the spectrum is presented.

\section{Spectrum at Weak Coupling Limit}
In Sec.~3, it is shown how the continuum spectrum of 1d model is recovered 
in the weak coupling limit. The spectrum by Eq.~(\ref{38}) is essentially $g^2k^2$, in which 
`$\,k\,$' represents the quanta of electric field in `$\, g\,$' units along the spatial direction. Recalling that 
in the present temporal gauge the momentum of the gauge 
field is related to the electric field by $\bm{E}=\bm{\dot{A}}$, and 
in absence of the magnetic field in the 1d model, the spectrum consists of only the kinetic term.
As we will see, the continuum spectrum expected by the classical model, like that of the 1d model in Sec.~3, 
can be recovered for higher dimensions as well. The main difference between 1d model and higher 
dimensional ones is that, the transfer-matrix of 1d model is diagonal and in the others it is block-diagonal 
\cite{kiafat,vadfat}. The matrix elements by the previous section show clearly that  
we are faced with almost equal values in each block, and finding the eigenvalues
of the transfer-matrix seems challenging. As the consequence, it is needed to go back to the field 
basis and manipulate the matrix elements to make a practically useful form to find the eigenvalues.
Back to (\ref{43}), we see that by 
\begin{align}\label{57}
\mathbb{P}=	\frac{1}{\sqrt{2}}\left(\begin{array}{c|c}
	\mathbb{1}_\mathrm{L} &  \mathbb{1}_\mathrm{L} \cr 
	\hline 
-\mathbb{1}_\mathrm{L}   &   	\mathbb{1}_\mathrm{L}
	\end{array}\right)
\end{align}
the matrix $\bm{C}$ comes to the form 
\begin{align}\label{58}
\bm{C'}= \mathbb{P}^{-1} \bm{C}\, \mathbb{P}= 	\left(\begin{array}{c|c}
2\,	\mathbb{1}_\mathrm{L}+\frac{1}{2}\bm{M}^T\!\bm{M} ~ & 0 \cr 
	\hline 
0   &   	\frac{1}{2}\bm{M}^T\!\bm{M}
	\end{array}\right)
\end{align}
By the explicit representations given in Appendix~A, it can be easily seen that the eigenvalues of 
$\bm{M}^T\!\bm{M}$ are non-negative. Accordingly, the zero eigenvalues of matrix $\bm{C}$ can
only happen in the lower right block. The number of these eigenvalues is presented in Tab~1. 
By the above representation, the matrix element (\ref{43}) takes the form (setting again $a=1$)
\begin{align}\label{59}
	\langle \bm{A'} |\widehat{V} | \bm{A}\rangle = \mathcal{A}\, 
\exp\!\left(-\frac{1}{2}\,\bm{A^{\!\mathsmaller{-}}} \Big(\mathbb{1}_\mathrm{L}+\frac{1}{4}\bm{M}^T\!\bm{M}\Big) 
\bm{A^{\!\mathsmaller{-}}} -\frac{1}{2}\,\bm{A^{\!\mathsmaller{+}}} \Big(\frac{1}{4}\bm{M}^T\!\bm{M}\Big) 
\bm{A^{\!\mathsmaller{+}}} \right)
\end{align}
in which 
\begin{align}\label{60}
\bm{A^{\!\mathsmaller{\pm}}}=\bm{A}\pm\bm{A'}
\end{align}
The above representation finds very simple form by going to the basis in which  
$\bm{M}^T\!\bm{M}$ is diagonal. As $\bm{M}^T\!\bm{M}$ is symmetric, there are orthonormal 
eigenvectors such that 
\begin{align}\label{61}
\bm{M}^T\!\bm{M}\, |\bm{\xi}\rangle = 4\xi^2\;|\bm{\xi}\rangle
\end{align}
for which $\langle\bm{\xi}| \bm{\xi'}\rangle=\delta_{\xi\xi'}$. The form of $4\xi^2$ for eigenvalues is chosen 
for later convenience. In this basis, obviously the zero eigenvalues only find contribution 
in the $\bm{A^{\!\mathsmaller{-}}}$ part, and the matrix element (\ref{59}) finds the form
\begin{align}
\langle \bm{A'} |\widehat{V} | \bm{A}\rangle = \mathcal{A}\, &
\exp\!\bigg(\!-\frac{1}{2}\sum_{\{\xi =0\}}(A_\xi-A'_\xi)^2  \cr
&-\frac{1}{2}\sum_{\xi\neq 0} \Big((1+\xi^2)(A_\xi-A'_\xi)^2 + \xi^2 (A_\xi+A'_\xi)^2\Big) \bigg)
\label{62}
\end{align}
in which $\{\xi=0\}$ is used to emphasize that the zero eigenvalue has degeneracy (see Tab.~1). 
As will be seen, the zero and non-zero modes correspond to the static and 
standing wave configurations, respectively. 
The absence of $A^{\!\mathsmaller{+}}_{\xi=0}$ in the above expression,
upon doing the Fourier transform, would result in the group volume 
$(2\pi/g)^{N_\mathrm{d}}$, as seen in the previous section.
The first term in the above is simply the free kinetic term for $A^{\!\mathsmaller{-}}_{\xi=0}$ modes. 
Both the group volume and the contribution by the free part to the spectrum can be recovered 
by the Fourier transform. In the $\xi$-basis, the Fourier term in (\ref{47}) takes the form
\begin{align}\label{63}
\mathrm{i}\,g\, \bm{\eta}^T\bm{\kappa} &= \mathrm{i}\,g\,(\bm{A}\cdot\bm{k} - \bm{A'}\cdot\bm{k'})\\
\label{64}
&=\frac{\mathrm{i}}{2}\,g \Big(\sum_{\{\xi =0\}}(A^{\!\mathsmaller{+}}_\xi k^{\!\mathsmaller{-}}_\xi 
+A^{\!\mathsmaller{-}}_\xi k^{\!\mathsmaller{+}}_\xi)+
\sum_{\xi \neq 0} (A^{\!\mathsmaller{+}}_\xi k^{\!\mathsmaller{-}}_\xi 
+A^{\!\mathsmaller{-}}_\xi k^{\!\mathsmaller{+}}_\xi) \Big)
\end{align}
in which $\bm{k^{\mathsmaller{\pm}}}=\bm{k}\pm\bm{k'}$.
As mentioned, the integration on $A^{\!\mathsmaller{+}}_{\xi=0}$ would develop 
the group volume, together with $\delta(k^{\!\mathsmaller{-}}_\xi)=\delta(k_\xi - k'_\xi)$ for $\xi=0$'s, 
which corresponds to $\delta(\bm{\tilde{\kappa}}_\mathrm{d})$ of the previous section.
The deltas are satisfied by the general requirement that $\bm{k}$ and 
$\bm{k'}$ belong to the same block, represented by (\ref{8}) and (\ref{10}),
$\bm{k}=\bm{k}_\ast + \qmb$ and $\bm{k'}=\bm{k}_\ast + \qmpb$.
The energy by the free modes can be calculated upon integration on $A^{\!\mathsmaller{-}}_{\xi=0}$'s
through the Fourier transform. As the consequence, the contribution of these modes to the 
energy, using $k_\xi=k'_\xi$ by the delta for the zero modes, is simply found to be 
\begin{align}\label{65}
E_{\mathrm{free}}= \sum_{\{\xi=0\}} \frac{1}{2}\left(\frac{g\, k^{\!\mathsmaller{+}}_\xi}{2}\right)^2
=\sum_{\{\xi=0\}} \frac{g^2 k^2_\xi}{2}
\end{align}
which is the free kinetic term on a circle of radius $1/g$. 
The allowed values for $\bm{k}$ by the Gauss law are discussed in Sec.~3. 
In the 1d case by Eq.~(\ref{38}), the free kinetic part represents the energy of 
the constant electric flux in the spatial direction. 
In the higher dimensional case the allowed electric fluxes may be a non-uniform one.
As an example is the straight flux $\bm{k}_\ast$ in Fig.~$6\,a$,
which represents a non-uniform flux as it is not repeated along the perpendicular 
direction, say $y$. Also, due to the term $\qmb$ in $\bm{k}=\bm{k}_\ast + \qmb$,
the original straight flux $\bm{k}_\ast$ may transform to a non-straight one, or may 
find an accompanied closed flux loop. Examples of both non-straight flux and a loop 
added one are presented in Fig.~$6\,b$ and $6\,c$, respectively. However, due to the 
projection to the subspace by the zero modes in (\ref{65}), it is only the uniform part
of the mentioned allowed states that contribute to the free part (\ref{65}). The projection to the
zero modes is simply done by the projection operator
\begin{align}\label{66}
P_0=\sum_{\{\xi=0\}} |\xi\rangle\langle\xi|
\end{align}
which satisfies $P_0^2=P_0$. By the explicit form of the matrix $\bm{M}$, it is a simple task 
to check that the result of the projection of the mentioned allowed states to the subspace by zero modes 
is a uniform electric flux, an example of which is presented in Fig.~11. The state in Fig.~11, by the convection 
introduced for labeling the links of a 2d lattice, is given by 
\begin{align}\label{67}
\bm{k}=(\undermat{N_s^2}{1,1,\cdots,}1,\undermat{N_s^2}{0,0,\cdots,}0)^T
\end{align}
\vskip .4cm
\noindent in which `$T$' is inserted to make it a column vector. 
Physically, the zero modes by the kinetic term $\bm{\dot{A}}^2$ are representing the 
pure electric field configurations, the stability of which forces them to be uniform. We later see that, 
how the continuum limit of the operator $\bm{M}^T\!\bm{M}$ only admits the uniform electric fields 
for the free kinetic part. The irrelevance of $\qmb$ to the zero mode sector can be understood easily 
by the action of the projection operator (\ref{66}). By the fact that the zero modes satisfy 
$\langle\xi|\bm{M}^T\!\bm{M}|\xi\rangle = \big|\bm{M}|\xi\rangle\big|^2 = 0$, we have 
\begin{align}\label{68}
P_0\, (\qmb)^T = \sum_{\{\xi=0\}} |\xi\rangle \langle\underbrace{\xi|\bm{M}}_0\!{}^T\!\!\cdot\! \bm{q}^T \! =0 
\end{align}
in which `$T$' again is inserted to make a column vector. 

\begin{figure}[H]
	\begin{center}
		\includegraphics[scale=.7]{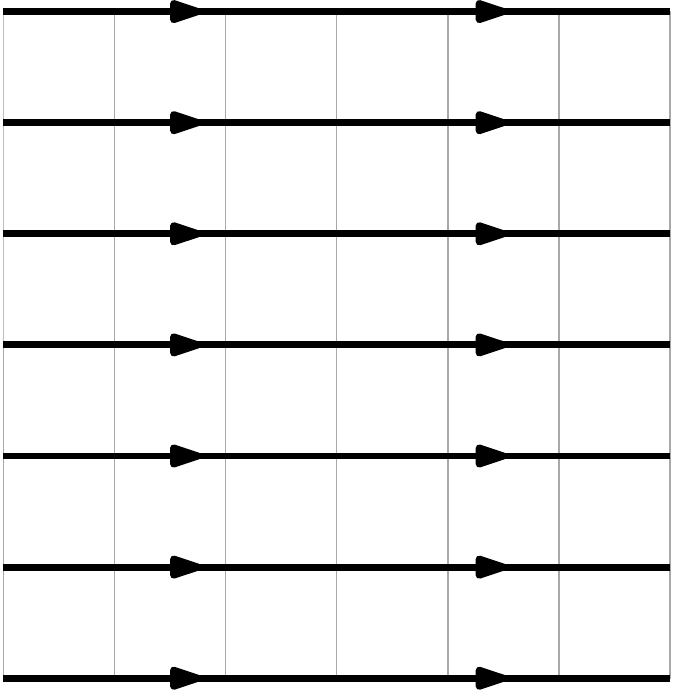}
%\vskip -.6cm   
\caption{\small 
The thick lines represent $\bm{k}$ as uniform electric fluxes on links of a 2d lattice.}
\label{fig11}
	\end{center}
\end{figure} \vskip -.6cm

The non-zero modes in the exponential of (\ref{62}) represent a harmonic oscillator dynamics, written
in the symmetric form in the potential term of the transfer-matrix element
\begin{align}\label{69}
\langle x' |\hat{V} | x\rangle =\sqrt{\frac{M}{2\pi}}\exp\bigg(\!\! -\frac{1}{2}M(x-x')^2 - \frac{1}{2}M\omega^2
\Big(\frac{x+x'}{2}\Big)^2\bigg)
\end{align}
with the spectrum $E_r=(r+\frac{1}{2})\,\omega$ by $r=0,1,2,\cdots$.
By the matrix element (\ref{62}) the frequencies are read as
\begin{align}\label{70}
\omega_\xi^2=\frac{4\xi^2}{1+\xi^2}
\end{align}
The harmonic oscillator nature of this part corresponds to the electromagnetic standing modes 
inside a resonance cavity. In particular, the electric and magnetic fields of the non-zero modes 
act as the kinetic and potential terms of an oscillator in the energy density $\frac{1}{2}(\bm{E}^2+\bm{B}^2)$. 
In the present model the term $\bm{A}\!\cdot\!\bm{M}^T\!\bm{M}\!\cdot\!\bm{A}$
is representing the magnetic part, and the kinetic term $(\bm{A}-\bm{A'})^2$
is for the electric part. By the explicit expression for the frequencies, we will
see that they in fact correspond to those in a resonance cavity.
By (\ref{70}), it is mentioned that the frequency has an upper limit (ultraviolet cutoff)
$\omega_\mathrm{max}^2= 4$. The existence of the cutoff
is simply a consequence of a lower limit of wavelength for a model on the lattice. 
All together, by adding up the contributions of zero and non-zero modes, 
after restoring the lattice parameter `$a$', the spectrum by the model is obtained
\begin{align}\label{71}
E_\mathrm{tot}=\sum_{\{\xi=0\}} \frac{g^2 k_\xi^2}{2\,a}+\frac{1}{a}
\sum_{\xi\neq 0} \big(r_\xi+\frac{1}{2}\big)\,\omega_\xi
\end{align}
Due to the $g^2k^2$ part, in the limit $g\ll1$, the above spectrum is a continuous one. 
This is equivalent to the case of a particle on a circle with radius $R$, with momenta and energy 
as $p=n/R$ and $E=n^2/(2mR^2)$, both being treated as continuous variables in the large $R$ limit. 
In the present model, the role of the radius is played in the field basis by $1/g$.

It is instructive to have more explicit expression for the frequency of standing modes. 
By the explicit expression presented in Appendix~A, for a 2d periodic lattice with $N_s$ sites 
in each direction, we have the following for the eigenvalues of $\bm{M}^T\!\bm{M}$ (see Appendix~B)
\begin{align}\label{72}
4\xi^2_{m,n}=4\left(\sin^2\frac{\pi m}{N_s}+\sin^2\frac{\pi n}{N_s}\right)
\end{align}
with $m, n=0,1,\cdots,N_s-1$. 
As mentioned earlier, the present derivations are valid for any lattice
size. In particular, the above eigenvalue and related frequency (\ref{70})
can be applied for arbitrary lattice size. This is an example of the announced 
feature that the formulation provides the possibility that different limits can be approached. 
In particular, in the large size limit $N_s\gg 1$, the frequencies can be approximated as
\begin{align}\label{73}
\omega^2_{m,n}\simeq 4\xi^2_{m,n}=\frac{4a^2\pi^2}{L^2}(m^2+n^2),~~~~~
m,n\ll N_s
\end{align}
in which $a\, N_s=L$ represents the size of the 2d square cavity. The above frequencies are
easily recognized as the allowed ones for standing waves in a box with periodic boundary conditions. 
By the representation given in Appendix~A for $\bm{M}^T\!\bm{M}$ in the continuum limit, 
the standing waves in a square of size $L$ satisfy
\begin{align}\label{74}
\begin{pmatrix} -\partial_y^2 & \partial_x \partial_y \cr 
\partial_x \partial_y & -\partial_x^2  \end{pmatrix} 
\begin{pmatrix} A(\vec{x},t)  \cr 
A'(\vec{x},t)  \end{pmatrix} = \partial_t^2 
\begin{pmatrix} A(\vec{x},t)  \cr 
A'(\vec{x},t)  \end{pmatrix}
\end{align}
in which 
\begin{align}\label{75}
A(\vec{x},t)=A_0\, e^{\frac{2\pi}{L}\mathrm{i}\,(m\,x+n\,y)-\mathrm{i}\,\omega\,t}\\
\label{76}
A'(\vec{x},t)=A'_0\, e^{\frac{2\pi}{L}\mathrm{i}\,(m\,x+n\,y)-\mathrm{i}\,\omega\,t}
\end{align}
The condition that for any amplitude $A_0$ and $A_0'$ there would be a solution 
as above leads to the frequency (\ref{73}). The other possibility, relevant to the 
zero mode sector with $\omega=0$, is the space independent solution
\begin{align}\label{77}
A(\vec{x},t)=A_1 \,t+ A_0\\
\label{78}
A'(\vec{x},t)=A'_1\, t+A'_0
\end{align}
which corresponds to the uniform electric field, as announced earlier. 
So in the large box limit the spectrum reads
\begin{align}\label{79}
E_\mathrm{tot}=\sum_{\{\xi=0\}} \frac{g^2 k_\xi^2}{2\,a}+\frac{2\pi}{L}
\sum_{m,n} \big(r_{m,n}+\frac{1}{2}\big)\,\sqrt{m^2+n^2}
\end{align}
The contribution by non-zero modes correspond to the radiation in a cavity. 
In the large $L$ limit, the discrete nature of the energy levels is relevant only for 
high frequency modes. 

\section{Conclusion and Summary}
The weak coupling limit of a lattice gauge model is commonly an obstacle to a 
full reconciliation between the lattice model and its counterpart on the continuum. 
In particular, the lattice gauge models are usually transferred 
to the extreme weak coupling regime in an uncontrolled way, leaving 
unresolved issues such as an unsought diverging group volume, as well as the unclear fate of  
main observable quantities in the lattice side, like the Wilson loops. 

In the previous sections, it was attempted to improve the procedure of taking
the weak coupling limit of a lattice gauge model. 
Based on the formulation of the transfer-matrix in the field Fourier basis, 
some issues raised by the weak coupling limit of the 
pure U(1) lattice gauge model in the temporal gauge were addressed. 
In \cite{vadfat,kiafat}, it was shown that the transfer-matrix in the Fourier basis is block diagonal.  
The members of each block can be constructed by a member of the block as the 
representative \cite{kiafat}. The matrix-element between two current-states of the same block 
is directly interpreted as the occurrence of all possible \textit{virtual link and loop currents} 
that transform the current-states to the vacuum. 

One of the basic tools used in the formulation of the transfer-matrix in the Fourier basis is 
the plaquette-link matrix $\bm{M}$ \cite{kiafat}, by which the fields and currents defined on the 
lattice can be managed in a completely controlled way at any coupling.
As a consequence, it is seen that the matrix $\bm{M}$ 
provides the possibility to keep and work with the fundamental lattice notions, 
such as links and sites, even in the extreme weak coupling limit.
On the other hand, by Sec.~4 and Appendix~A, using this matrix enables to translate the tools on 
the continuum into the lattice side; examples are
the correspondence (\ref{87}) between operations $\bm{M} \longleftrightarrow 
(-\partial_y ~~ \partial_x)$, and relations (\ref{90}) and (\ref{91}).
Similar correspondences are presented for 3d lattice as well.
These all make it possible to calculate the dimension of the subspace by zero eigenvalues 
of the operator and to handle the group volume in a safe way, as seen previously.  
Based on notions and expressions developed in \cite{kiafat}, as far as the transfer-matrix 
of the U(1) gauge model in the Fourier basis is concerned, the following clarifications were made:
\begin{enumerate}
\item In the Fourier basis the gauge invariant states identified by the Gauss law constraint 
consist of Fourier states with equal link-current and without boundary. These states, which 
are known as Wilson loops in the field basis, are either 
current-loops belonging to the vacuum block or states with 
equal links-currents along periodic or infinite spatial directions. In the strong 
coupling limit, the current-loops of the vacuum block have the main contribution to the transfer-matrix
as well as the spectrum, and link-currents along the spatial directions cost an infinite energy.
In the weak coupling regime, however, the link-currents along spatial directions also find comparable roles. 
\item In the extreme weak coupling limit, the matrix in the quadratic action 
is identified as the origin of the diverging contributions to the elements of the transfer-matrix in the 
Fourier basis. The states belonging to the subspace corresponding to the zero eigenvalues are clearly 
interpreted as pure gauge configurations, on which the matrix in the quadratic action vanishes. The 
dimension of the subspace as well as the diverging volume of the subspace in the weak coupling limit can be 
handled and extracted in a safe and controlled way.
\item The spectrum by the lattice model is obtained analytically at the 
weak coupling limit. The calculation is by means of the very basic notions and tools of the lattice model
for any dimension and size of lattice. The obtained spectrum consists of the contribution by 
the static and standing wave field configurations on the lattice.
The spectrum at the weak coupling limit corresponds to the expected one 
by the continuum model in the large lattice limit.
\end{enumerate}

\begin{appendices}

\section{Matrices $\bm{M}$ and $\bm{C}$ for 2d and 3d Models}

In this Appendix the considerations about the dimension of the sub-space by pure gauge 
configurations, leading to zero eigenvalues of the matrix $\bm{C}$, is presented. Also, the 
related notions and elements in both lattice and the continuum models are derived and compared. 
The following is done for 2d model first, and the very same construction for 3d case is presented
shortly. In the temporal gauge $A_0\equiv 0$, in which 
\begin{align}\label{80}
\bm{E}=\bm{\dot{A}},~~~~ \bm{B}=\bm{\nabla} \times \bm{A}
\end{align}
 the Hamiltonian density 
\begin{align}\label{81}
\mathcal{H}=\frac{1}{2}
\left(\bm{\dot{A}}^2+(\bm{\nabla} \times \bm{A})^2\right)
\end{align}
in 2d takes the form
\begin{align}\label{82}
\mathcal{H}=\frac{1}{2}
\left(\dot{A}_x^2+\dot{A}_y^2+(\partial_x A_y-\partial_y A_x)^2\right)
\end{align}
By the two-adjacent times interpretation of (\ref{5}) for the definition of the transfer-matrix, the Hamiltonian
symmetrized between the variables $A$ and $A'$ at two times, after integration by parts, is
\begin{align}\label{83}
\mathcal{H}=\frac{1}{2}&
\Big((A'_x-A_x)^2+(A'_y-A_y)^2\cr
&+\frac{1}{2}\big[(A_x\partial_y-A_y\partial_x)(\partial_x A_y-\partial_y A_x)
+A\to A' \big]\Big)
\end{align}
By the definition for the vector $\bm{\eta}$ as (\ref{42}), the Hamiltonian density recasts to
\begin{align}\label{84}
\mathcal{H}=\frac{1}{2}\, 
\bm{\eta}^T \bm{C}_\mathrm{2d}\, \bm{\eta}
\end{align}
in which
\begin{align}\label{85}
\bm{C}_\mathrm{2d}=
\begin{pmatrix} 1 & -1 \cr 
-1 & 1  \end{pmatrix}
\otimes \mathbb{1}_{\vec{x}} + \frac{1}{2} \, \mathbb{1}_\mathrm{2} \otimes 
\begin{pmatrix} -\partial_y^2 & \partial_x \partial_y \cr 
\partial_x \partial_y & -\partial_x^2  \end{pmatrix}
\end{align}
Comparing the above operator with (\ref{45}) of the lattice formulation, we see that 
the combination $\bm{M}^T\!\bm{M}$ is in fact acting as the derivative $\partial_i$ in
the last $2\times 2$ matrix:
\begin{align}\label{86}
\bm{M}^T\!\bm{M} \longleftrightarrow 
\begin{pmatrix} -\partial_y^2 & \partial_x \partial_y \cr 
\partial_x \partial_y & -\partial_x^2  \end{pmatrix} = 
\begin{pmatrix} \partial_y \cr -\partial_x \end{pmatrix}  
\begin{pmatrix} -\partial_y ~~~ \partial_x \end{pmatrix}  
\end{align} 
which, recalling that $\partial_i$ is an antisymmetric operator ($\partial_i^T=-\partial_i$), 
leads to the following correspondence
\begin{align}\label{87}
\bm{M} \longleftrightarrow 
\begin{pmatrix} -\partial_y ~~~ \partial_x \end{pmatrix}  
\end{align} 
The matrix $\bm{M}$ can be defined by using the $N_s\times N_s$ translation-matrix 
$\bm{T}$, defined by its elements \cite{vadfat}
\begin{align}\label{88}
T_{ab}= \delta_{ab}-\delta_{a+1,b}-\delta_{a,N_s}\,\delta_{b1},~~~~~
a,b=1,\cdots,N_s
\end{align}
Then, the general form of the matrix $\bm{M}$ for the 2d lattice is given in \cite{vadfat,kiafat} 
\begin{align}\label{89}
\bm{M}=\left(\begin{array}{c|c}
  & \cr
~-\bm{M}_y~~~
& ~~~~ \bm{M}_x ~~~ \cr 
&  
\end{array}\right)
\end{align}
by which, using (\ref{87}), we have
\begin{align}
\label{90}
\bm{M}_x = - \bm{T}\otimes \mathbb{1}_{N_s} & 
~\longleftrightarrow \ \partial_x \\
\label{91}
\bm{M}_y = -\mathbb{1}_{N_s}\otimes \bm{T}&
~\longleftrightarrow \ \partial_y  
\end{align}
By construction, the matrix $\bm{M}$ is $N_s^2\times 2N_s^2$ dimensional, as it should be for the 2d lattice.
The similarity between the lattice formulation in the weak coupling limit and the continuum model 
in fact goes further, as both operators vanish acting on 
a gradient. For the above operator the gradient of function is
\begin{align}\label{92}
\bm{\nabla}f=\begin{pmatrix} \partial_x f \cr 
\partial_y f  \end{pmatrix}
\end{align}
and obviously satisfies
\begin{align}\label{93}
\begin{pmatrix} -\partial_y ~~~ \partial_x \end{pmatrix}  
\begin{pmatrix} \partial_x f \cr \partial_y f \end{pmatrix} =0 
\end{align}
In the lattice side, the gradient of a function is simply given by (\ref{90}) and (\ref{91}).  
Let us use the notation that the function `$\, f\,$' at site `$\, i\,$' is denoted as $f_i$.
Then the column vector is defined as:
\begin{align}\label{94}
\vec{f} = (f_1,f_2,\cdots,f_{N^{~^{\!\!\! 2}}_{\! s}} )^T
\end{align}
in which `$T$' is inserted to make it a column vector.
Using (\ref{90}) and (\ref{91}), the gradient on lattice is then defined as
\begin{align}\label{95}
\bm{\Delta} f=\begin{pmatrix} \Delta_x f \cr 
\Delta_y f  \end{pmatrix}
=\begin{pmatrix} \bm{M}_x\, \vec{f} \cr 
\bm{M}_y \, \vec{f}  \end{pmatrix}
\end{align}
which, using $\bm{M}_x \bm{M}_y = \bm{M}_y\bm{M}_x$
by the given representations, satisfies 
\begin{align}\label{96}
\bm{M}\cdot\begin{pmatrix}\Delta_x f \cr \Delta_y f \end{pmatrix}=
\begin{pmatrix} -\bm{M}_y ~~ \bm{M}_x \end{pmatrix} 
\begin{pmatrix} \bm{M}_x\, \vec{f} \cr 
\bm{M}_y \, \vec{f}  \end{pmatrix}  =0 
\end{align}

\begin{figure}[t]%[!ht]
	\begin{center}
		\includegraphics[scale=.7]{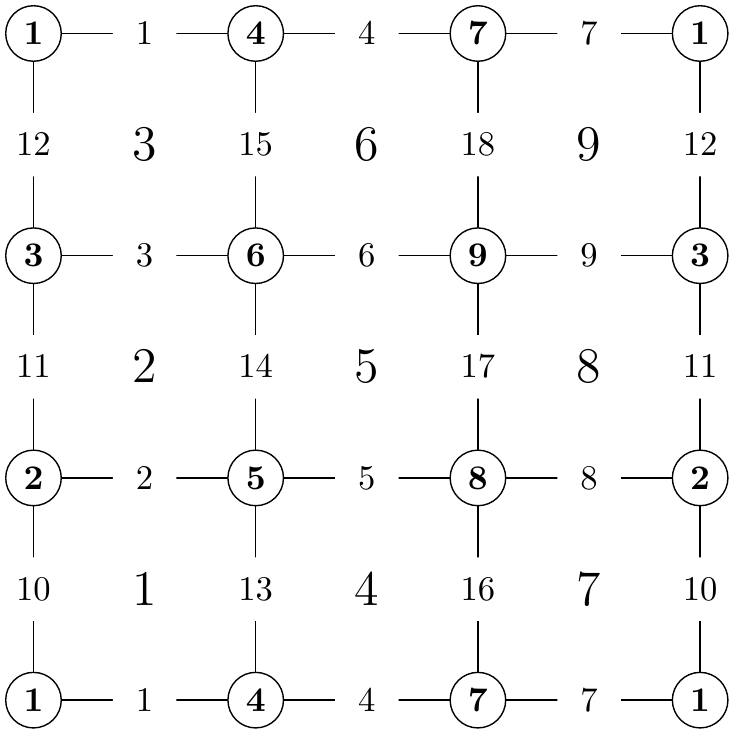}
%\vskip -.6cm   
\caption{\small The numbering of links and plaquettes for the $3\times 3$ 2d 
\textit{periodic} lattice used in (\ref{19}) as the representation of matrix $\bm{M}$ \cite{vadfat}. }
\label{fig12}
	\end{center}
\end{figure} \vskip -.2cm

It is useful to have an explicit representation for the plaquette-link matrix
$\bm{M}$. For the site, plaquette and link numberings of the $3\times 3$ periodic lattice given 
in Fig.~\ref{fig12}, using the definition (\ref{3}), one finds the following form for the $9\times 18$ dimensional 
matrix $\bm{M}$ \cite{vadfat}
\begin{align}\label{97}
\bm{M}=\left(\begin{array}{c@{\shrink}c@{\shrink}c@{\shrink}c@{\shrink}
c@{\shrink}c@{\shrink}c@{\shrink}c@{\shrink}c@{\shrink\shrink}
c@{\shrink}c@{\shrink}c@{\shrink}c@{\shrink}c@{\shrink}c@{\shrink}
c@{\shrink}c@{\shrink}c}
\msp  & \msm  & \msz  & \msz  & \msz  & \msz  & \msz  & \msz  & \msz  & \msm  & \msz  & \msz  & \msp  & \msz  & \msz  & \msz  & \msz  & \msz  \\ [-8pt]
 \msz  & \msp  & \msm  & \msz  & \msz  & \msz  & \msz  & \msz  & \msz  & \msz  & \msm  & \msz  & \msz  & \msp  & \msz  & \msz  & \msz  & \msz  \\ [-8pt]
 \msm  & \msz  & \msp  & \msz  & \msz  & \msz  & \msz  & \msz  & \msz  & \msz  & \msz  & \msm  & \msz  & \msz  & \msp  & \msz  & \msz  & \msz  \\ [-8pt]
 \msz  & \msz  & \msz  & \msp  & \msm  & \msz  & \msz  & \msz  & \msz  & \msz  & \msz  & \msz  & \msm  & \msz  & \msz  & \msp  & \msz  & \msz  \\ [-8pt]
 \msz  & \msz  & \msz  & \msz  & \msp  & \msm  & \msz  & \msz  & \msz  & \msz  & \msz  & \msz  & \msz  & \msm  & \msz  & \msz  & \msp  & \msz  \\ [-8pt]
 \msz  & \msz  & \msz  & \msm  & \msz  & \msp  & \msz  & \msz  & \msz  & \msz  & \msz  & \msz  & \msz  & \msz  & \msm  & \msz  & \msz  & \msp  \\ [-8pt]
 \msz  & \msz  & \msz  & \msz  & \msz  & \msz  & \msp  & \msm  & \msz  & \msp  & \msz  & \msz  & \msz  & \msz  & \msz  & \msm  & \msz  & \msz  \\ [-8pt]
 \msz  & \msz  & \msz  & \msz  & \msz  & \msz  & \msz  & \msp  & \msm  & \msz  & \msp  & \msz  & \msz  & \msz  & \msz  & \msz  & \msm  & \msz  \\ [-8pt]
\undermat{-\bm{M}_y}
{ \msz  & \msz  & \msz  & \msz  & \msz  & \msz  & }\msm  & \msz  &  \msp  & 
\undermat{\bm{M}_x}
{ \msz  & \msz  & \msp  & \msz  & \msz  & \msz  & } \msz  &  \msz  & \msm   \\
\end{array}
\right)
\end{align}\vskip .8cm
By the representation, the gradient based on the given ordering of 
sites and links for the $3\times 3$ periodic lattice, denoting $f_{ij}=f_j-f_i$, is the following
\begin{align}\label{98}
\Delta f =\begin{pmatrix} \bm{M}_x\, \vec{f} \cr 
\bm{M}_y \, \vec{f}  \end{pmatrix}
= (&f_{41},f_{52},f_{63},f_{74},f_{85},f_{96},f_{17},f_{28},f_{39}, \cr
&f_{21},f_{32},f_{13},f_{54},f_{65},f_{46},f_{87},f_{98},f_{79})^T
\end{align}
We notice that, for example, $f_{63}$ is sitting in place of the variable on link `3', 
consistent by the numbering given in Fig.~12.

In the language of gauge theories, the expressions (\ref{92}) and (\ref{95}) are referred as pure gauge configurations.
In fact, in the present temporal gauge $A_0\equiv 0$, starting with zero field configurations
in two adjacent times given by $\bm{A}=\bm{A'}=0$, the remaining spatial gauge transformations 
leads to $\bm{A}=\bm{A'}=\bm{\nabla} f$ in the continuum theory, 
and $\bm{A}=\bm{A'}=\bm{\Delta} f$ in the lattice one. As expected, by the pure gauge 
configurations (\ref{92}) and (\ref{95}), the Hamiltonian vanishes, leading to vanishing 
of the strength fields; $\bm{E}=\bm{B}=\bm{0}$. 

It is instructive to see that the construction for the 2d model can be directly generalized to the 3d one.
By the two-adjacent times interpretation in the definition of the transfer-matrix, the Hamiltonian
symmetrized between the variables $A$ and $A'$ at two times, after integration by parts, takes the 
following form in 3d
\begin{align}\label{99}
\mathcal{H}=&\frac{1}{2}\Big((A'_x-A_x)^2+(A'_y-A_y)^2+(A'_z-A_z)^2
\cr&+\frac{1}{2}\Big[(A_x\partial_y-A_y\partial_x)(\partial_x A_y-\partial_y A_x)
+A\to A'\cr&~~~~+(A_y\partial_z-A_z\partial_y)(\partial_y A_z-\partial_z A_y)
+A\to A'\cr&~~~~+(A_z\partial_x-A_x\partial_z)(\partial_z A_x-\partial_x A_z)
+A\to A'\Big]\Big)
\end{align}
written in matrix form
\begin{align}\label{100}
\mathcal{H}=\frac{1}{2}\, 
\bm{\eta}^T \bm{C}_\mathrm{3d}\, \bm{\eta}
\end{align}
in which 
\begin{align}\label{101}
\bm{C}_\mathrm{3d}=
\begin{pmatrix} 1 & -1 \cr 
-1 & 1  \end{pmatrix}
\otimes \mathbb{1}_{\vec{x}} + \frac{1}{2} \, \mathbb{1}_\mathrm{2} \otimes 
\begin{pmatrix} -\partial_y^2-\partial_z^2 & \partial_x \partial_y & \partial_x \partial_z  \cr 
\partial_x \partial_y & -\partial_x^2  -\partial_z^2& \partial_z \partial_y \cr 
\partial_x \partial_z & \partial_z \partial_y & -\partial_x^2  -\partial_y^2\end{pmatrix}
\end{align}
Comparing with (\ref{44}), 
\begin{align}\label{102}
\bm{M}^T\!\bm{M} \longleftrightarrow &
\begin{pmatrix} -\partial_y^2-\partial_z^2 & \partial_x \partial_y & \partial_x \partial_z  \cr 
\partial_x \partial_y & -\partial_x^2  -\partial_z^2& \partial_z \partial_y \cr 
\partial_x \partial_z & \partial_z \partial_y & -\partial_x^2  -\partial_y^2\end{pmatrix} \\
\label{103}
= &\begin{pmatrix} \partial_y  & \partial_z & 0\cr
-\partial_x & 0  & \partial_z \cr
0 & -\partial_x & -\partial_y\end{pmatrix}  
\begin{pmatrix} -\partial_y & \partial_x & 0\cr
-\partial_z & 0 & \partial_x \cr
0 & -\partial_z & \partial_y\end{pmatrix}  
\end{align} 
For a periodic 3d lattice with $N_s$ sites in each direction we have for the number of links and plaquettes 
\begin{align}\label{104}
\nl=\np=3N_s^3
\end{align}
suggesting 
\begin{align}\label{105}
\bm{M}=\left(\begin{array}{c|c|c} 
-\bm{M}_y & \bm{M}_x & 0 \cr
\hline
-\bm{M}_z & 0 & \bm{M}_x\cr
\hline
0 & -\bm{M}_z & \bm{M}_y
\end{array}\right) \longleftrightarrow 
\begin{pmatrix} -\partial_y & \partial_x & 0\cr
-\partial_z & 0 & \partial_x \cr
0 & -\partial_z & \partial_y\end{pmatrix}  
\end{align} 
with
\begin{align}\label{106}
\bm{M}_x &= -\bm{T} \otimes\mathbb{1}_{N_s}\otimes\mathbb{1}_{N_s}~ \longleftrightarrow ~\partial_x \\
\label{107}
\bm{M}_y &= -\mathbb{1}_{N_s}\otimes \bm{T} \otimes\mathbb{1}_{N_s} ~\longleftrightarrow~ \partial_y\\
\label{108}
\bm{M}_z &= -\mathbb{1}_{N_s} \otimes\mathbb{1}_{N_s}\otimes  \bm{T}~ \longleftrightarrow~ \partial_z
\end{align}
by which as it is required $\bm{M}_i\bm{M}_j=\bm{M}_j\bm{M}_i$ for $i,j=x,y,z$.

\section{Eigenvalues of $\bm{M}^T\!\bm{M}$}

The eigenvalues of $\bm{M}^T\!\bm{M}$ for periodic 2d lattice are calculated in this part. The
3d case can be calculated straightforwardly. By the given representation, $\bm{M}^T\!\bm{M}$
for 2d lattice comes to the form
\begin{align}\label{109}
\bm{M}^T\!\bm{M} = 
\left(\begin{array}{c|c}
\mathbb{1}\otimes \bm{T}^\dagger\bm{T}~ & -\bm{T}\otimes \bm{T}^\dagger  \cr 
	\hline 
-\bm{T}^\dagger\otimes \bm{T}   &   	\bm{T}^\dagger\bm{T}\otimes \mathbb{1}
	\end{array}\right)
\end{align}
in which both $\mathbb{1}$ and $\bm{T}$ are $N_s\times N_s$ dimensional (as in Appendix~A).
So $\bm{M}^T\!\bm{M}$ is $2N_s^2\times 2N_s^2$, as it should. 
For a block matrix 
\begin{align}\label{110}
M=\left(\begin{array}{c|c}
A & B \cr 
	\hline 
C   & D
	\end{array}\right)
\end{align}
if $C$ and $D$ commute, then $\det M = \det (A\,D-B\,C)$.
The calculation is done in the basis that $\bm{T}$ is diagonal, in which all blocks commute.
The secular equation $\det(\bm{M}^T\!\bm{M}-\lambda)=0$ then finds the form
\begin{align}\label{111}
\det\Big(\lambda\big(\lambda \, \mathbb{1}\otimes \mathbb{1} - \mathbb{1}\otimes \bm{T}^\dagger\bm{T}    
-\bm{T}^\dagger\bm{T} \otimes \mathbb{1}  \big)\Big) =0 
\end{align}
The outer $\lambda$ leads to $N_s^2$ number of zeros as eigenvalues. 
Other eigenvalues, denoting eigenvalues of $\bm{T}^\dagger\bm{T}$ as
$\beta_m$, are simply
\begin{align}\label{112}
\lambda_{m,n} = \beta_m+\beta_n,~~~m,n=0,1,\cdots,N_s-1
\end{align}
The only remaining part is to find the eigenvalues of $\bm{T}^\dagger\bm{T}$. By the explicit
representation given in Appendix~A, the secular equation for $\bm{T}$ is
\begin{align}\label{113}
(1-\alpha)^{N_s}=1
\end{align}
with the following as a solution
\begin{align}\label{114}
\alpha_m=1-\exp\frac{2\pi\mathrm{i}\,m}{N_s}
\end{align}
So the eigenvalues of $\bm{T}^\dagger\bm{T}$ are 
\begin{align}\label{115}
\beta_m=\alpha_m \alpha^*_m=4\sin^2\frac{\pi m}{N_s}
\end{align}
resulting in 
\begin{align}\label{116}
\lambda_{m,n} =4\left(\sin^2\frac{\pi m}{N_s}+\sin^2\frac{\pi n}{N_s}\right) ,~~~m,n=0,1,\cdots,N_s-1
\end{align}

\end{appendices}

\vskip .8cm
\textbf{Acknowledgment:}
A.H.F. is grateful to Dr. N. Vadood for providing an earlier explicit representation of  
matrix $\bm{M}$ for the 3d lattice. 
The helpful comments by M. Khorrami on the manuscript are gratefully acknowledged. 
The authors also would like to thank the anonymous referee for providing detailed and helpful 
comments leading to correct and improve this work. 
This work is supported by the Research Council of Alzahra University.

\end{document}